\documentclass[10pt,journal,compsoc]{IEEEtran}

\usepackage{url}
\usepackage{amsmath,amsfonts}
\usepackage[linesnumbered,ruled]{algorithm2e}
\usepackage{graphicx}
\usepackage{textcomp}
\usepackage{xcolor}
\usepackage{ tipa }
\usepackage{tabularx}
\usepackage{url}
\usepackage{color}
\usepackage{soul}
\usepackage{caption}
\usepackage{subcaption}
\usepackage{mwe}
\usepackage{enumerate}
\usepackage[utf8]{inputenc}
\usepackage[english]{babel}
\usepackage{amsmath}
\usepackage{amsthm}
\usepackage{mathtools}
\usepackage{csquotes}
\usepackage{blkarray, bigstrut}
\usepackage{gauss}
\usepackage[tikz]{bclogo}
\usepackage{lipsum}
\usepackage{booktabs}
\usepackage{cite}
\usepackage{amsmath,amssymb,amsfonts}
\usepackage{algorithmic}
\usepackage{graphicx}
\usepackage{textcomp}
\usepackage{xcolor}
\usepackage{colortbl} 
\usepackage{makecell}
\usepackage[hidelinks]{hyperref}
\usepackage[most]{tcolorbox}

\usepackage{cuted}

\definecolor{coWith}{HTML}{35739E}
\definecolor{coWithout}{HTML}{DB8132}
\newcommand{\swatch}[1]{\fcolorbox{black!55}{#1}{\rule{0pt}{1.1ex}\rule{1.1ex}{0pt}}}

\newcommand{\summarybox}[3]{%
  \vspace{4mm}
  \noindent\framebox[\columnwidth][c]{%
    \parbox[b]{0.95\columnwidth}{%
      \textit{\textbf{#1.} #2} \\[2mm] 
      #3 
    }
  }
  \vspace{4mm}
}

\tcbset{textmarker/.style={
        enhanced,
        parbox=true,boxrule=0mm,boxsep=1mm,arc=0mm,
        outer arc=1mm,left=2mm,right=0mm,top=7pt,bottom=7pt,
        toptitle=1mm,bottomtitle=1mm,oversize}}

\newtcolorbox{hintBox}{textmarker,
    borderline west={6pt}{0pt}{yellow},
    colback=yellow!10!white}
\newtcolorbox{importantBox}{textmarker,
    borderline west={6pt}{0pt}{red},
    colback=red!10!white}
\newtcolorbox{noteBox}{textmarker,
    borderline west={8pt}{0pt}{gray},
    colback=gray!10!white}

\theoremstyle{definition}

\theoremstyle{remark}

\usetikzlibrary{arrows.meta}

\usepackage{tikz}
\usetikzlibrary{plotmarks}
\usetikzlibrary{arrows,shapes,positioning}
\usetikzlibrary{decorations.markings}
\tikzstyle arrowstyle=[scale=1]
\tikzset{>=latex}

\usepackage[colorinlistoftodos]{todonotes}

\definecolor{chestnut}{rgb}{0.8, 0.36, 0.36}

\definecolor{chestnut}{rgb}{0.8, 0.36, 0.36}

\definecolor{revOneColor}{HTML}{1A53A0}   
\definecolor{revTwoColor}{HTML}{1B7837}   
\definecolor{revThreeColor}{HTML}{7B2D8E} 

\newif\ifshowrevmarks
\showrevmarksfalse   

\definecolor{revOneR2Color}{HTML}{C8500A}   
\definecolor{revThreeR2Color}{HTML}{0B7A75} 
\definecolor{revAER2Color}{HTML}{555555}    
\newif\ifshowrevtwomarks
\showrevtwomarksfalse
\makeatletter
\newcommand{\revb@mark}[4]{
  \ifshowrevtwomarks
    {\color{#1}#4}%
    \def\revb@lbl{#3}%
    \ifx\revb@lbl\@empty
      \textsuperscript{\textcolor{#1}{\tiny[#2]}}%
    \else
      \textsuperscript{\textcolor{#1}{\tiny[#2:#3]}}%
    \fi
  \else
    #4%
  \fi}
\newcommand{\rtwoone}[2][]{\revb@mark{revOneR2Color}{R1r2}{#1}{#2}}
\newcommand{\rtwothree}[2][]{\revb@mark{revThreeR2Color}{R3r2}{#1}{#2}}
\newcommand{\rtwoae}[2][]{\revb@mark{revAER2Color}{AEr2}{#1}{#2}}
\makeatother

\makeatletter
\newcommand{\rev@mark}[4]{
  \ifshowrevmarks
    {\color{#1}#4}%
    \def\rev@lbl{#3}%
    \ifx\rev@lbl\@empty\else
      \textsuperscript{\textcolor{#1}{\tiny[#2:#3]}}%
    \fi
  \else
    #4%
  \fi}
\newcommand{\revone}[2][]{\rev@mark{revOneColor}{R1}{#1}{#2}}
\newcommand{\revtwo}[2][]{\rev@mark{revTwoColor}{R2}{#1}{#2}}
\newcommand{\revthree}[2][]{\rev@mark{revThreeColor}{R3}{#1}{#2}}
\makeatother

\AtBeginDocument{%
  \providecommand\BibTeX{{%
    \normalfont B\kern-0.5em{\scshape i\kern-0.25em b}\kern-0.8em\TeX}}}

\begin{document}
\title{Automated Code Review Assignments: \\ An Alternative Perspective of \\ Code Ownership on GitHub}

\author{Jai~Lal~Lulla, Raula~Gaikovina~Kula, Christoph~Treude
\IEEEcompsocitemizethanks{
\IEEEcompsocthanksitem Raula is with The University of Osaka, Japan.\newline  E-mail: raula-k@ist.osaka-u.ac.jp.
\IEEEcompsocthanksitem Jai and Christoph are with Singapore Management University, Singapore.\newline  E-mail: ctreude@smu.edu.sg
}
}

\IEEEtitleabstractindextext{%
\begin{abstract}
Code ownership is central to ensuring accountability and maintaining quality in large-scale software development. Yet, as external threats such as software supply chain attacks on project health and quality assurance increase, mechanisms for assigning and enforcing responsibility have become increasingly critical. In 2017, GitHub introduced the CODEOWNERS feature, a file in the repository that names the developers responsible for specific files and automatically requests them as reviewers, to strengthen accountability and protect critical parts of the codebase. Despite its potential, little is known about how CODEOWNERS is actually adopted and practiced.
We present the first large-scale empirical study of CODEOWNERS usage across over 844,000 pull requests with 1.9 million comments and over 2 million reviews.
We identify 10,287 designated owners, the developers named in CODEOWNERS files, track their review activities, and compare them with contribution-based owners, the developers who contributed at least 5\% of a file's lines.
\rtwoone[2]{Results indicate that designated owners take part in a moderate share of the pull requests that touch their files,} that their collaborative behavior resembles that of contribution-based owners, \rtwoone[1]{and that after adoption merge time and review effort grow more slowly, or fall, relative to their pre-adoption trend.}
Using regression discontinuity design analysis, we further find that \rtwoone[4]{designated owners' review, decision (approval and merge), and authoring activity jumps upward at adoption, and that over the following year they do more review and decision work per pull request than before, whereas contribution-based owners do less review work per pull request.} Our results position CODEOWNERS as a promising yet \revtwo[S2]{sparsely adopted} mechanism for improving software governance and resilience. We discuss how projects can leverage this alternative ownership method as a perspective to enhance \rtwoone[1]{security and accountability} in open-source development.

\end{abstract}

}

\maketitle

\IEEEdisplaynontitleabstractindextext

\IEEEpeerreviewmaketitle

\section{Introduction}
\label{sec:intro}

In the increasingly interconnected ecosystem of open-source software, effective code ownership and maintenance practices are essential to ensure quality, security, and accountability. With the popularity of collaborative platforms like GitHub, large-scale projects now involve contributions from developers worldwide, necessitating robust mechanisms for managing contributions and code reviews. To address these challenges, GitHub introduced the CODEOWNERS feature in 2017,\footnote{\url{https://github.blog/news-insights/product-news/introducing-code-owners/}}
 which allows repository administrators to assign specific users or teams as owners of particular files or \revthree[1:41]{entities}. By automatically requesting reviews from designated owners whenever relevant files are modified,\footnote{\url{https://shorturl.at/dQYxq}}
 CODEOWNERS aims to strengthen accountability and streamline the review process for critical parts of a codebase.

Existing research has shown that code ownership and management practices significantly affect software quality, security, and maintainability~\cite{bird2011don,Avelino2019}. Traditionally, ownership has been defined by contribution thresholds—developers who authored a certain percentage of a file (e.g., 5\% or 10\%) were considered “owners”~\cite{Rahman2011}. Following Foucault et al.~\cite{foucalEtAll2014}, we adopt a similar approach and define a developer as a contribution-based owner of a file if they have contributed at least 5\% of its total lines of code. Such contribution-based ownership models have been linked to lower defect rates, faster issue resolution, and overall better project health. Other studies have shown correlations between ownership practices and metrics such as code coverage, commit quality, and team productivity~\cite{Kola-Olawuyi2024,Thongtanunam2016}.

However, no prior study has examined ownership as explicitly specified in configuration files—a form of ownership policy that defines responsibility independent of contribution history. Nordberg~\cite{Nordberg2003} argues that ownership encompasses multiple models. \revthree[MAJ1]{Explicit ownership is, moreover, not limited to CODEOWNERS: it can also be expressed through other means---for example \texttt{OWNERS} or \texttt{MAINTAINERS} files, service-ownership tooling, or ownership simply recorded in project documentation or spreadsheets---so CODEOWNERS represents one platform-native point in a broader design space rather than the sole alternative to contribution-based ownership.} We posit that GitHub’s CODEOWNERS feature introduces a distinct and underexplored form of explicit ownership with direct implications for both software reliability and security. As software supply chain threats and dependency risks grow, understanding how such formal ownership mechanisms are adopted and enforced becomes critical for safeguarding open-source ecosystems. While prior research inferred ownership indirectly through contribution-based metrics (e.g., commits or file modifications), the implications of explicitly declared ownership remain unexplored. \revone[1.I]{Even where the repositories we study overlap with those examined in earlier contribution-based work, our subject of analysis differs in kind: we examine the explicitly \emph{declared} ownership policy itself, rather than ownership inferred post hoc from commit counts or line-level contribution shares.} \revone[1.II]{We study GitHub rather than another platform for both practical and methodological reasons. CODEOWNERS originated on GitHub in 2017, and comparable features were only later added by other platforms such as GitLab;\footnote{\url{https://docs.gitlab.com/ee/user/project/codeowners/}} concentrating on the platform where the feature originated therefore lets us observe its use over a longer period than would be possible on platforms that adopted it afterwards. Methodologically, restricting the study to a single platform holds constant platform-level factors like the review interface, branch-protection semantics, and default behaviors that would otherwise confound a cross-platform comparison, while GitHub exposes the pull-request, review, and \texttt{git blame} data our analysis requires through a uniform API. We do not claim that GitHub is representative of all hosts; generalization to other platforms is treated as a threat to external validity (Section~\ref{sec:threats}).} Our study is the first to empirically examine how declared ownership—through GitHub's CODEOWNERS—shapes collaboration, review behavior, and workload.

\revone[1.II]{Our aim in studying CODEOWNERS is practical, not only descriptive. By measuring who is designated as an owner, how completely a repository is covered, and how review behavior shifts after adoption, we give maintainers and project managers evidence for deciding whether, where, and how to declare ownership. Such a decision currently rests on little empirical footing: a team weighing the upkeep of a CODEOWNERS file against its benefits has had few numbers to consult. Our findings on adherence (RQ1), on the gap between declared owners and top contributors (RQ2), \rtwoone[4]{and on how review workload changes after adoption (RQ3 and RQ4)} speak to that decision and to how review responsibilities can be allocated across a team.}

The CODEOWNERS file, located in the root, \texttt{.github}, or \texttt{docs} directories, leverages a simple pattern-based syntax to define which users or teams are responsible for specific parts of a repository. Ownership declarations can be granular, covering individual files, directories, or wildcard patterns. For example, \texttt{docs/*} assigns ownership of all files within the \texttt{docs} directory, while \texttt{docs/**} covers all subdirectories recursively. CODEOWNERS files work in conjunction with GitHub's branch protection rules, which can require code owner approval before merges, ensuring that designated reviewers assess critical changes. Multiple owners or teams can share responsibility, enabling collaborative review while reinforcing governance and stability.

\begin{figure*}
\centering
\includegraphics[height=1.3334\columnwidth]{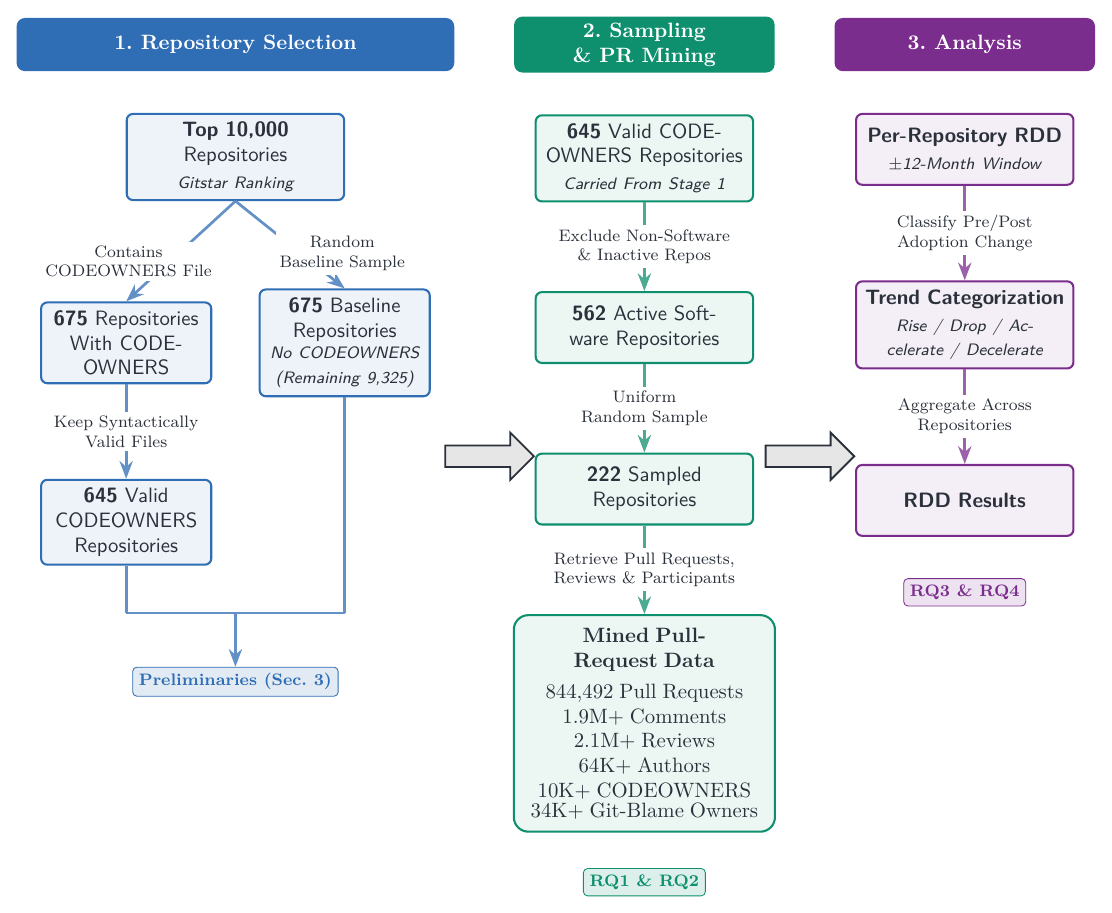}
\caption{Overview of our study showing data collection, preparation, and designation to each RQ.}
\label{fig:data_pipeline}
\end{figure*}

In this study, we leverage GitHub's CODEOWNERS feature to offer a new perspective on code ownership in the context of pull requests and code review.
Only about 7\% of top-starred GitHub repositories currently use CODEOWNERS, yet its adoption concentrates on critical non-code assets such as build workflows, dependency manifests, and licensing files—suggesting a governance and security orientation rather than simple code review automation. By systematically examining usage patterns across these 645 repositories, we conducted a large scale quantitative analysis of the review process addressing four research questions:
\begin{description}
\item[RQ1] \textbf{To what extent do GitHub projects adhere to CODEOWNERS?}
\item[RQ2] \textbf{To what extent do CODEOWNER roles differ from contribution-based ownership metrics?}
\item[RQ3] \textbf{What is the impact of CODEOWNERS on the dynamics of the pull request process?}
\item[RQ4] \textbf{What is the impact of CODEOWNERS on developers' review workloads?}
\end{description}

\rtwoone[2]{Results indicate that designated owners take part in the pull requests that touch their files, but only in a moderate and far-from-universal share of them (RQ1).}
We find that CODEOWNERS often differ from contribution-based owners, indicating a deliberate separation between authorship and responsibility (RQ2).
\rtwoone[1]{Using regression discontinuity analysis, we observe that after adoption the trends in merge time, comment volume, and reviewer participation decelerate, growing more slowly or falling relative to their pre-adoption course (RQ3).}
\rtwoone[4]{Finally, designated owners' review, decision (approval and merge), and authoring activity jumps upward at adoption, and over the following year they do more review and decision work per pull request than before, whereas contribution-based owners do less review work per pull request (RQ4).}

This work represents the first comprehensive empirical analysis of explicit, configuration-based code ownership on GitHub. By studying the adoption, coverage, and behavioral effects of CODEOWNERS, we bridge a gap between contribution-based ownership models and modern, policy-driven governance mechanisms. Our findings reveal that while CODEOWNERS adoption remains limited, \rtwoone[4]{it concentrates on protecting critical assets, raises the review and decision activity of designated owners, and may reduce the risk of unreviewed or unsafe changes.}
Our data and scripts are publicly available on 
\href{https://zenodo.org/records/17533176?preview=1&token=eyJhbGciOiJIUzUxMiJ9.eyJpZCI6ImY4ZDMzNmY4LTc1ZDYtNGNjZi1iN2UwLThlMGRmODI3M2E1YyIsImRhdGEiOnt9LCJyYW5kb20iOiIwNGRjMTkwNzhmYTFlZjE3NzBkNTNmN2FjM2RhNWQyYSJ9.tqgKgeSfsZRw2rk3iEQhs539ldr9rnQm0fCKtBWF2PR25kyC002sKyBeLBaMOYiH_UMxJIxbB8SDoTnfEI1l-w}{Zenodo}.
\section{Dataset}

This section outlines our study design, including repository selection, data collection, sampling, and resulting dataset characteristics. An overview of the data acquisition pipeline is provided in Figure~\ref{fig:data_pipeline}.

\subsection{Repository Selection}

To capture representative patterns of code ownership across popular projects, we began with the 10,000 most-starred GitHub repositories. Stars are a widely used indicator of project visibility and engagement, making this sample suitable for studying mainstream development practices.

Due to GitHub API query limitations,\footnote{\url{https://docs.github.com/en/rest?apiVersion=2022-11-28}}
 repository metadata was obtained from Gitstar Ranking,\footnote{\url{https://gitstar-ranking.com/repositories}}
 a reliable aggregation of GitHub's public data. We developed a Selenium-based web scraper\footnote{\url{https://selenium-python.readthedocs.io/}}
 to collect repository URLs efficiently.

Next, we identified repositories that explicitly used the CODEOWNERS feature by searching, via the GitHub API, for files named CODEOWNERS in recognized locations (root, \texttt{.github/}, or \texttt{docs/}). This filtering yielded 675 repositories. For comparison, a baseline sample of 675 repositories was randomly selected from the remaining 9,325 without CODEOWNERS files, enabling contrasts between projects with and without explicit ownership specifications.

Each CODEOWNERS repository was then downloaded and analyzed using the CODEOWNERS Parser tool.\footnote{\url{https://github.com/hmarr/codeowners}}
 Of the 675 repositories, 645 contained syntactically valid CODEOWNERS files, which formed our initial dataset of projects explicitly employing code ownership policies.

\subsection{Data Collection Procedures}

For both repository groups (with and without CODEOWNERS), we collected a wide range of project-level metrics—issue count, pull request count, commit count, contributor count, primary language, and repository age—to capture development activity and project characteristics.
Most data were retrieved through the GitHub REST API, while contributor counts were scraped using Selenium to ensure accuracy, as API values occasionally include unfiltered contributors that are excluded on the web interface due to GitHub's internal deduplication filters.\footnote{\url{https://github.com/orgs/community/discussions/24355}} We treated contributor counts from the web interface as ground truth.

To ensure temporal consistency, all repositories were cloned and analyzed using a fixed cutoff date of 9 October 2024. \revone[2.III]{We fixed a single cutoff because mining data at this scale spanned an extended collection period. Without a common cutoff, repositories downloaded later in the process would have accumulated more recent pull-request activity than those downloaded earlier, producing a sliding-window effect in which the most recently collected repositories contribute disproportionately more recent data and could unduly influence the observed trends. Anchoring every repository to the same cutoff date places them on a common temporal footing and removes this collection-induced bias. We discuss the recency of this cutoff, and its potential impact on the conclusions, as a threat to validity in Section~\ref{sec:threats}.} We implemented a custom API wrapper on top of the \texttt{requests} library\footnote{\url{https://pypi.org/project/requests/}} to handle batching, retries, and caching, thereby improving throughput and minimizing token consumption under GitHub's rate limits.

\subsection{Sampling Strategy}

The complete set of 645 valid repositories containing CODEOWNERS files was not used directly for pull request (PR)-level analysis.
Because large-scale PR retrieval is resource-intensive and subject to GitHub API limitations, we analyzed a representative subset of repositories rather than the full population.
Before sampling, we performed a filter to ensure only relevant, active software projects were included:

\begin{itemize}
\item \revone[2.I]{We excluded non-software repositories (e.g., demos, tutorials, or documentation-only projects) through manual inspection, recording each repository's classification in a spreadsheet.}
\item \revone[2.I]{We excluded repositories with no commit or pull-request activity in the 12~months preceding data collection, and archived or disabled repositories.}
\end{itemize}

This filtering yielded 562 active, software-oriented repositories.
From these, we drew a \revone[2.II]{simple random sample of 222 repositories, selected uniformly at random from the 562 filtered repositories, for detailed PR-level analysis.
Technical constraints such as GitHub API rate limits and available access tokens influenced the overall sample size but not the random selection procedure.}

A sample size of 192 would have provided a 90\% confidence level with a 5\% margin of error; our final sample of 222 achieves approximately 4.5\% margin of error, thus maintaining statistical robustness.
\revone[2.II]{Unless otherwise stated, all pull-request-level statistics (Table~\ref{tab:dataset_stats}) and RQ1-RQ4 analyses are based on this curated subset of 222 repositories. The preliminary characterization in Section~\ref{sec:prelim} instead uses the full sets: the 645 valid-CODEOWNERS repositories and the 675 baseline repositories, which were selected only by the presence or absence of a valid CODEOWNERS file and otherwise remain the unrestricted top-starred repositories. As preliminaries rely solely on repository-level metadata, they did not require the in-depth, per-pull-request mining (reviews, comments, or \texttt{git blame} ownership) needed for the other research questions. Because RQ3 and RQ4 analyses estimate each repository's change around its own CODEOWNERS adoption event, they have no counterpart in the baseline repositories, which never adopt CODEOWNERS and provide no before/after cutoff to measure; we therefore mine pull requests only for adopting repositories.}

\subsection{Dataset Characteristics}

Table~\ref{tab:dataset_stats} summarizes the resulting dataset. In total, we analyzed over 840,000 pull requests and \revthree[3:56]{nearly 2.1 million reviews}, capturing a comprehensive view of how CODEOWNERS interacts with developer activity and collaboration.
This dataset provides a rich empirical foundation for investigating explicit code ownership, enabling quantitative comparisons between traditional contribution-based ownership and configuration-defined CODEOWNERS practices.

\begin{table}
\centering
\caption{Summary statistics of the Dataset collected for the sample of Repositories from the 645 Repositories}
\label{tab:dataset_stats}
\begin{tabular}{lr}
\toprule
\textbf{Metric} & \textbf{Value} \\
\midrule
Total pull requests collected & 844,492 \\
Total comments & 1,915,525 \\
Total reviews & 2,096,964 \\
Total review comments & 1,771,573 \\
Contribution-based owners identified & 34,701 \\
CODEOWNERS identified & 10,287 \\
Unique authors & 64,230 \\
\bottomrule
\end{tabular}
\end{table}

\section{Preliminaries on the 645 Repositories}
\label{sec:prelim}

Before addressing the main research questions, we first provide an overview of how the CODEOWNERS feature is used across popular GitHub repositories. This preliminary analysis characterizes its adoption, repository contexts, rule structures, and the roles of declared owners, providing essential context for interpreting later results.

\subsection{Adoption of CODEOWNERS Files}

We began by quantifying the adoption of GitHub's CODEOWNERS feature among high-profile repositories. Among the 10,000 most-starred repositories, 675 (6.75\%) included a CODEOWNERS file in one of the recognized locations. Of these, 645 files (96\%) were valid and successfully parsed, while the remainder contained syntax or path-related errors.

This relatively low adoption rate indicates that explicit code ownership remains a selective practice even among prominent projects, suggesting that many continue to rely on informal or contribution-based ownership mechanisms.

\subsection{Repository Characteristics}

\begin{figure*}
\centering
\begin{subfigure}[b]{0.33\textwidth}
\centering
\includegraphics[width=\linewidth]{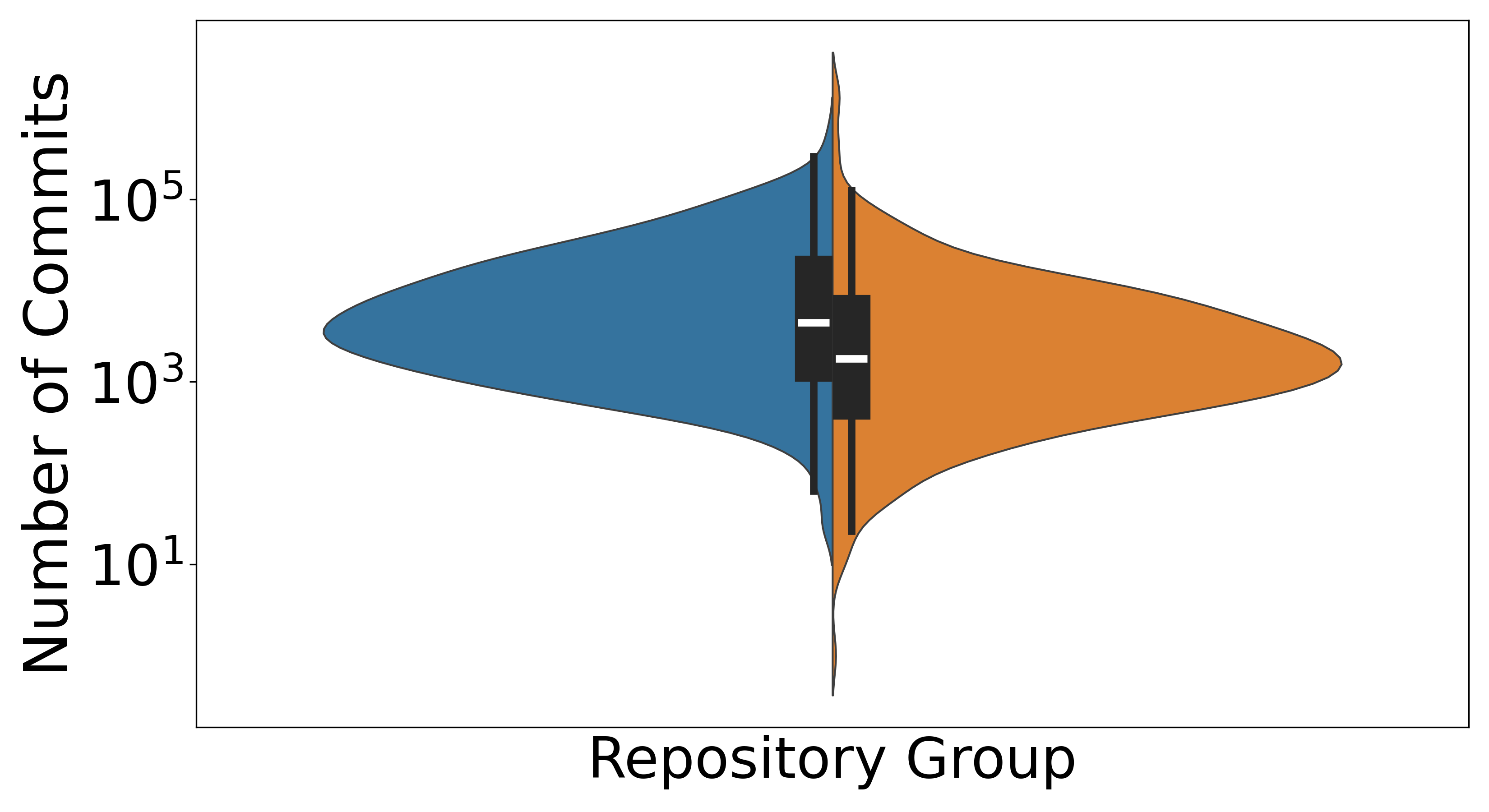}
\caption{Commits}
\label{fig:commits_count_distribution}
\end{subfigure}\hfill
\begin{subfigure}[b]{0.33\textwidth}
\centering
\includegraphics[width=\linewidth]{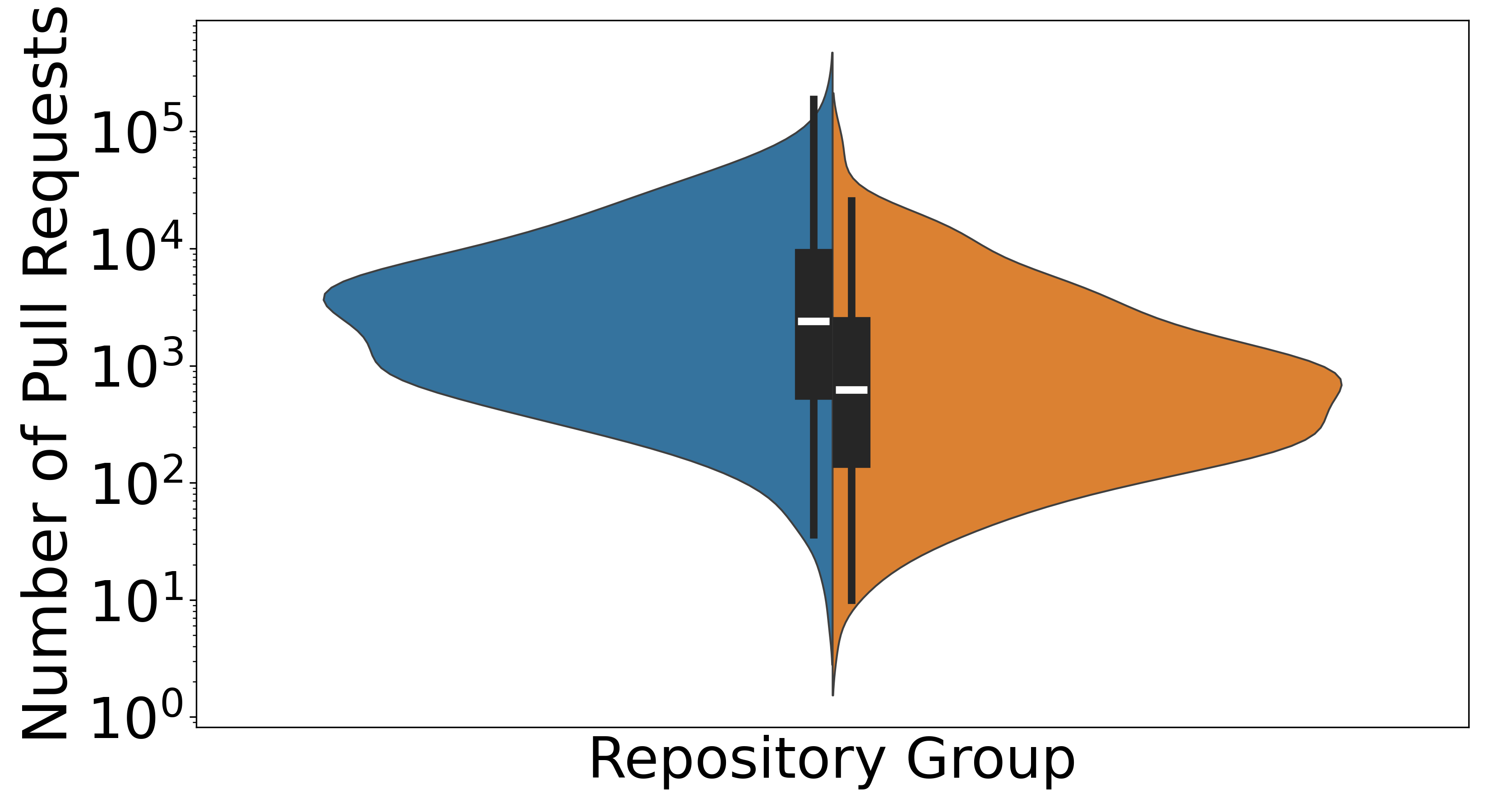}
\caption{Pull requests}
\label{fig:pull_requests_count_distribution}
\end{subfigure}\hfill
\begin{subfigure}[b]{0.33\textwidth}
\centering
\includegraphics[width=\linewidth]{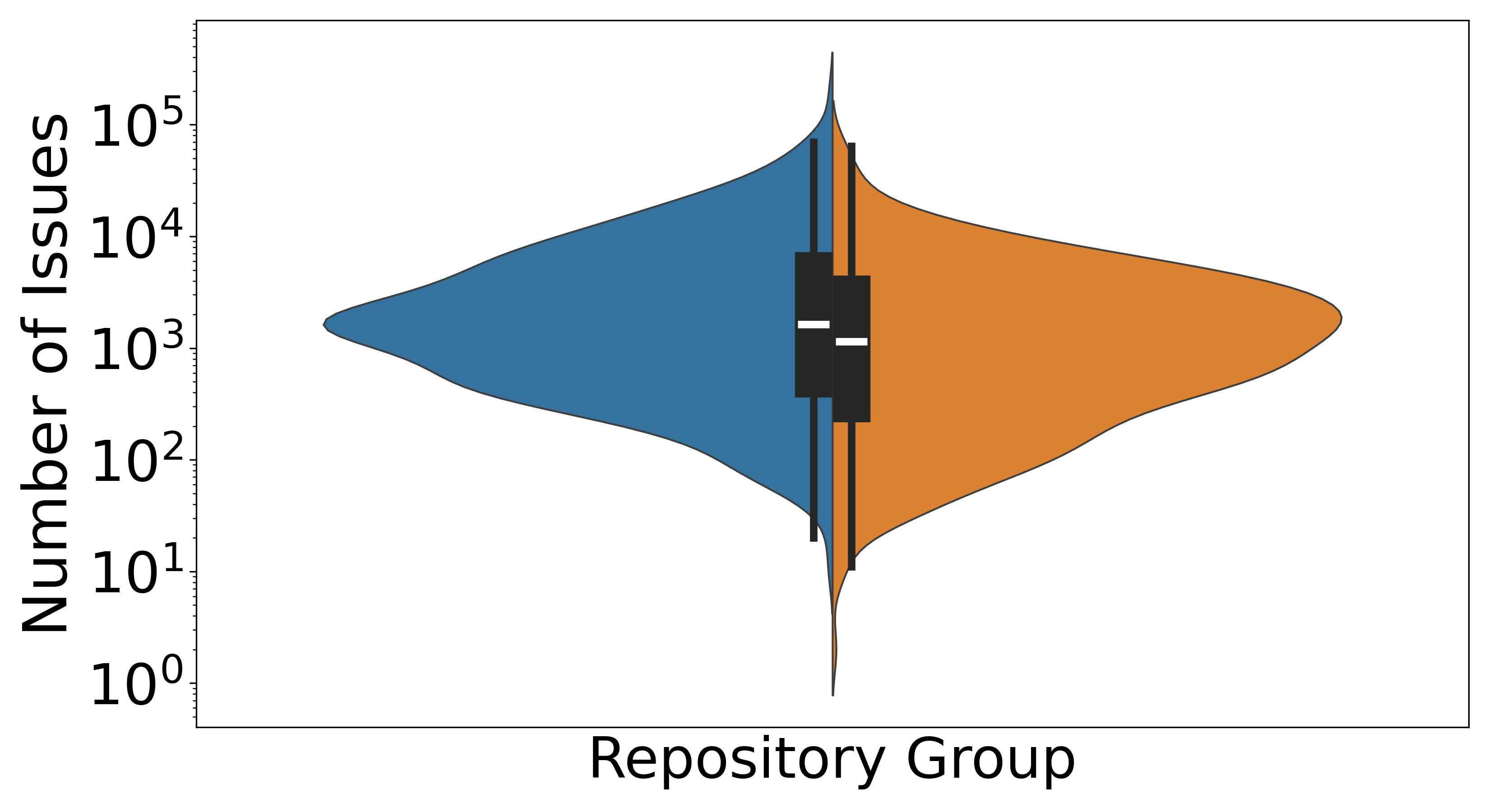}
\caption{Issues}
\label{fig:issues_count_distribution}
\end{subfigure}

\vspace{2mm}

\begin{subfigure}[b]{0.33\textwidth}
\centering
\includegraphics[width=\linewidth]{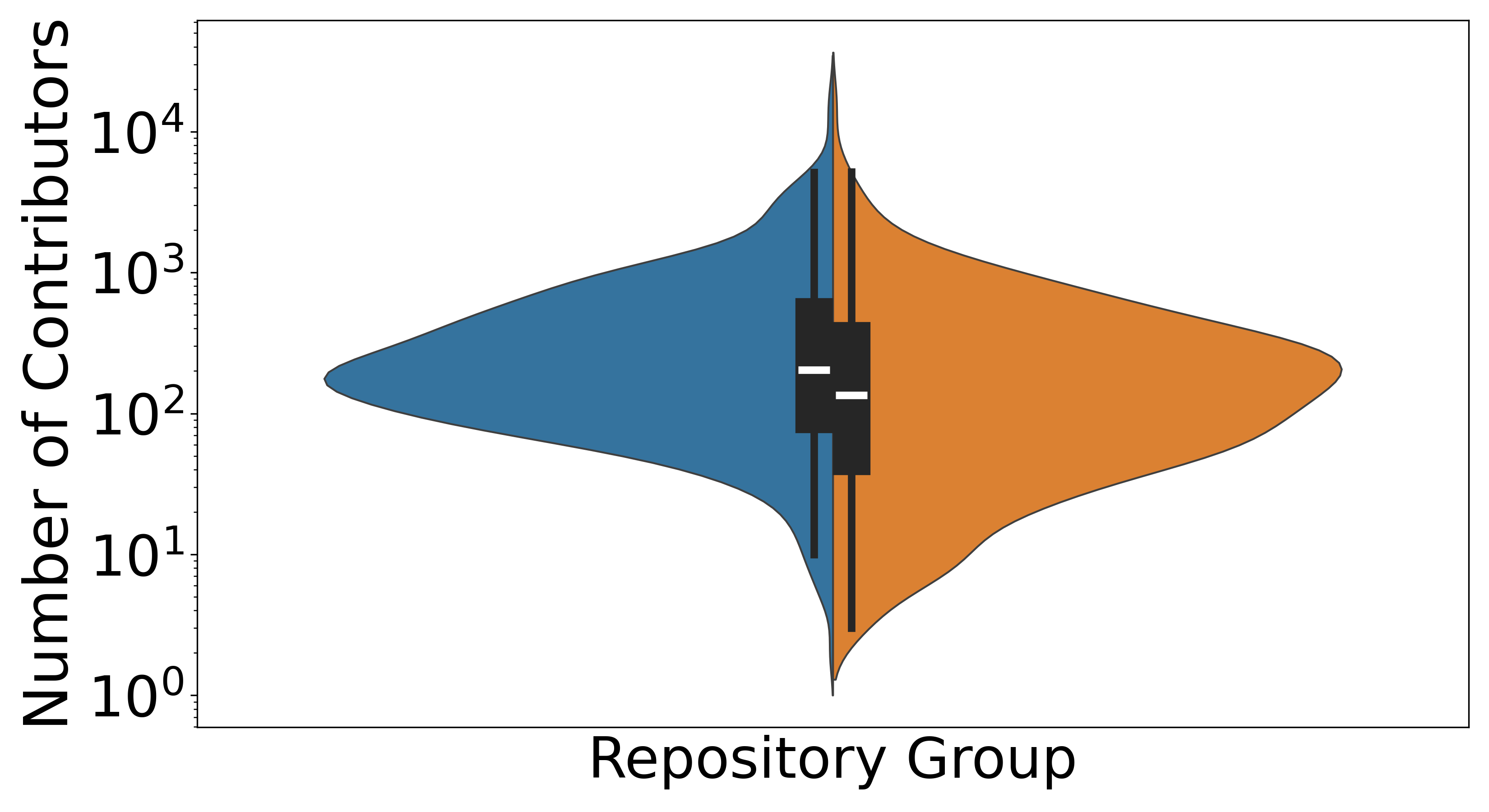}
\caption{Contributors}
\label{fig:contributors_count_distribution}
\end{subfigure}\hspace{0.02\textwidth}
\begin{subfigure}[b]{0.33\textwidth}
\centering
\includegraphics[width=\linewidth]{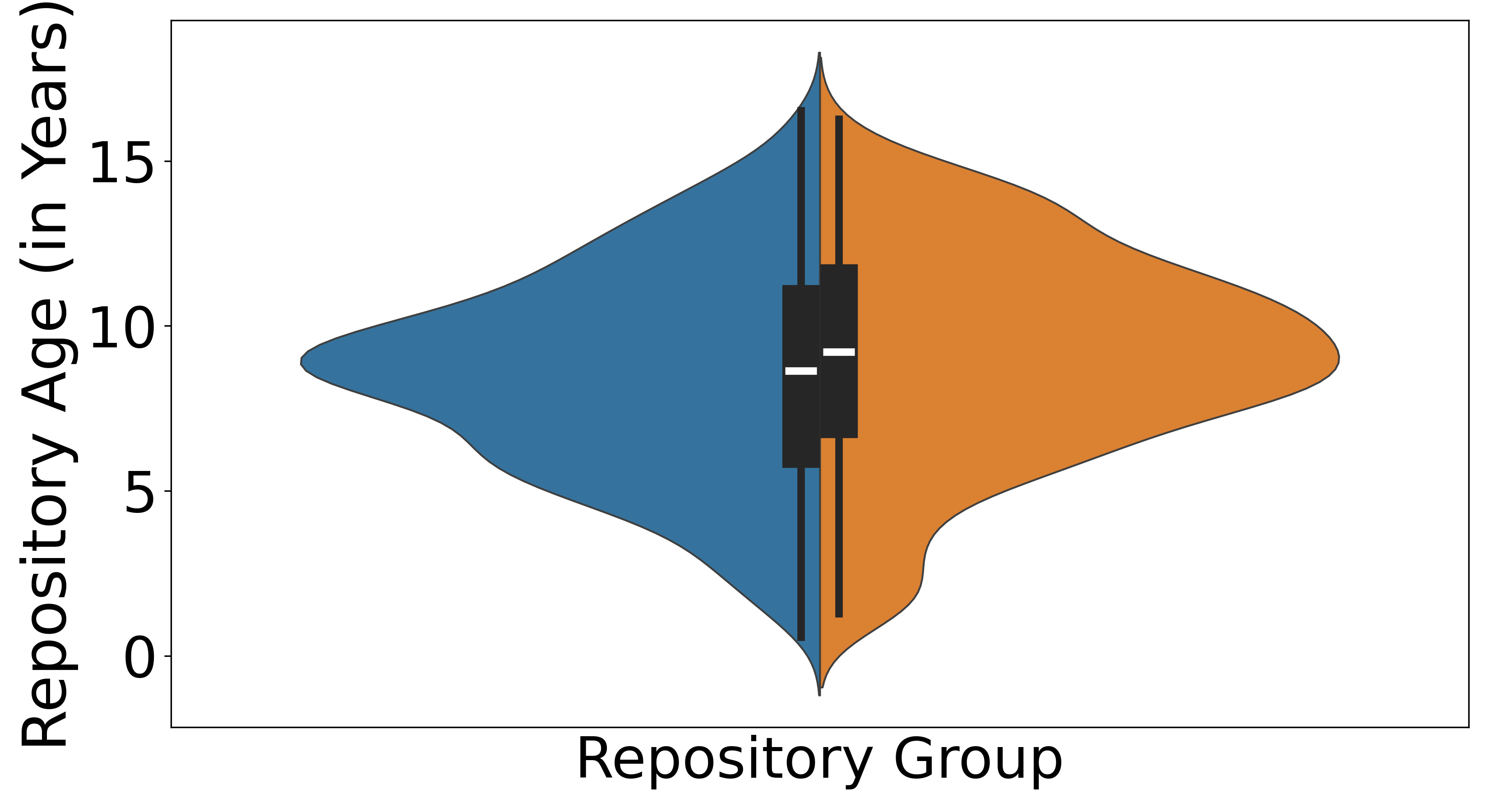}
\caption{Repository age}
\label{fig:repo_age_distribution2}
\end{subfigure}

\caption{Distributions of commits, pull requests, issues, contributors, and repository age for repositories with and without CODEOWNERS files. Each violin is split by group: \swatch{coWith}~with CODEOWNERS (left half) and \swatch{coWithout}~without CODEOWNERS (right half). The inner box and whiskers mark the median, interquartile range, and $1.5\times$IQR. Panels (a)--(d) use a logarithmic $y$-axis; panel (e) uses a linear scale in years.}
\label{fig:repo-characteristics}
\end{figure*}

To understand the kinds of projects that adopt CODEOWNERS, we compared repository-level metrics between those with and without such files. Figure~\ref{fig:repo-characteristics} illustrates statistically significant differences ($p < 0.05$, Mann-Whitney U and Kolmogorov-Smirnov tests).
Repositories with CODEOWNERS files consistently showed higher activity levels—more commits, pull requests, issues, and contributors—than those without. Specifically, the median number of commits was 4,488 vs. 1,789, pull requests 2,394 vs. 621, and issues 1,632 vs. 1,149. These repositories also tended to be slightly younger (median age 8.64 years vs. 9.21) and involved more contributors (median 204 vs. 135).

This pattern suggests that CODEOWNERS adoption is particularly prevalent in active, collaborative projects, where structured review and accountability mechanisms are most beneficial.
\revtwo[S2]{This characterization is limited to repository-level activity and size. We do not analyze the problem domain or application area of adopters and therefore cannot determine whether projects of a particular kind, such as security-sensitive, infrastructure, or dependency-heavy ones, are more likely to adopt CODEOWNERS. Identifying domain-level drivers would require content analysis of the repositories, which we leave to future work.}

\subsection{Structure and Coverage of CODEOWNERS Rules}

\begin{figure}
\centering
\begin{subfigure}[b]{0.5\textwidth}
\centering
\includegraphics[width=0.9\columnwidth]{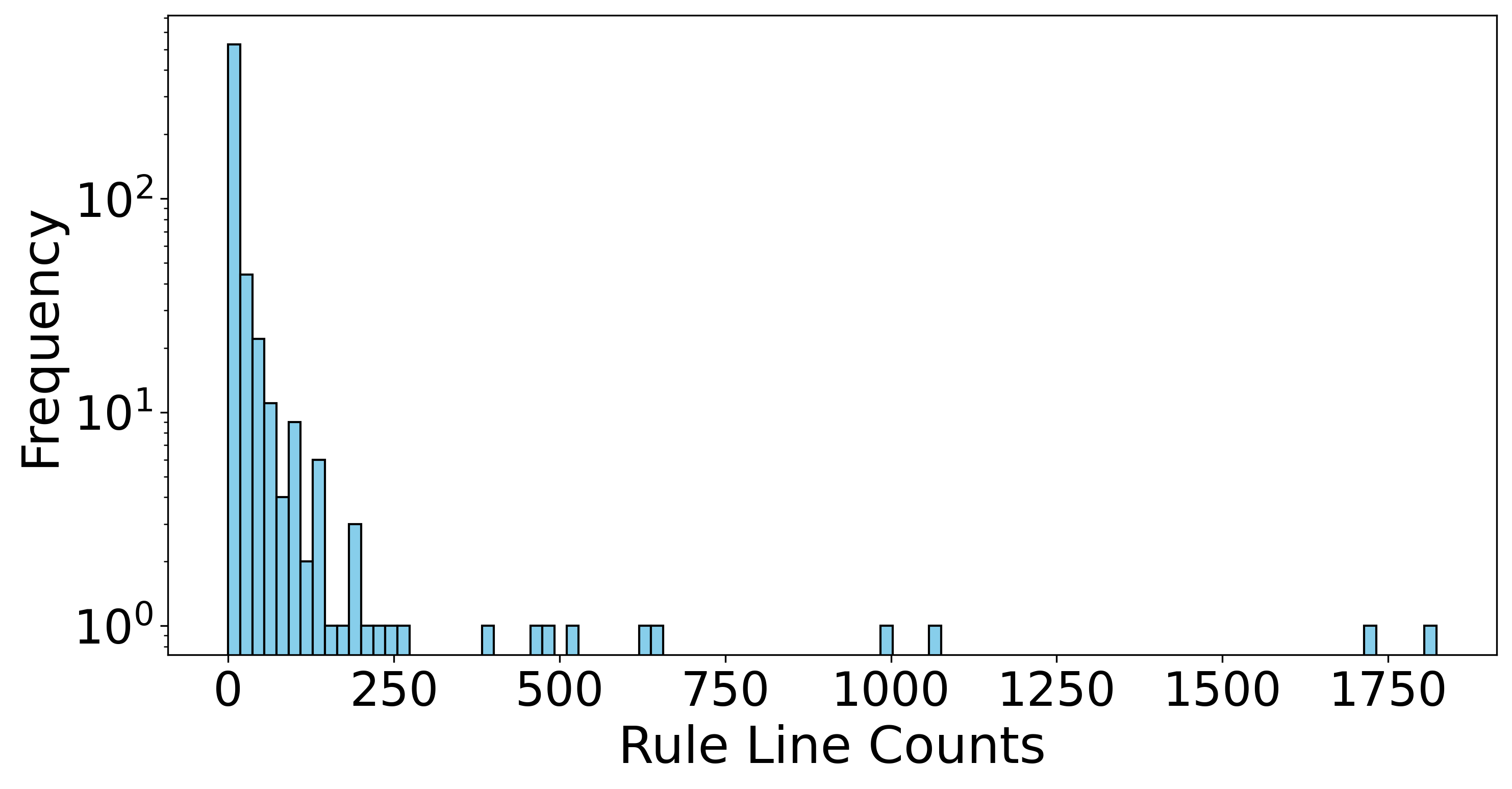}
\caption{Distribution of Rule Line Counts in CODEOWNERS Files (one outlier omitted).}
\label{fig:rule_line_counts_distribution}
\end{subfigure}
\begin{subfigure}[b]{0.5\textwidth}
\centering
\includegraphics[width=0.9\columnwidth]{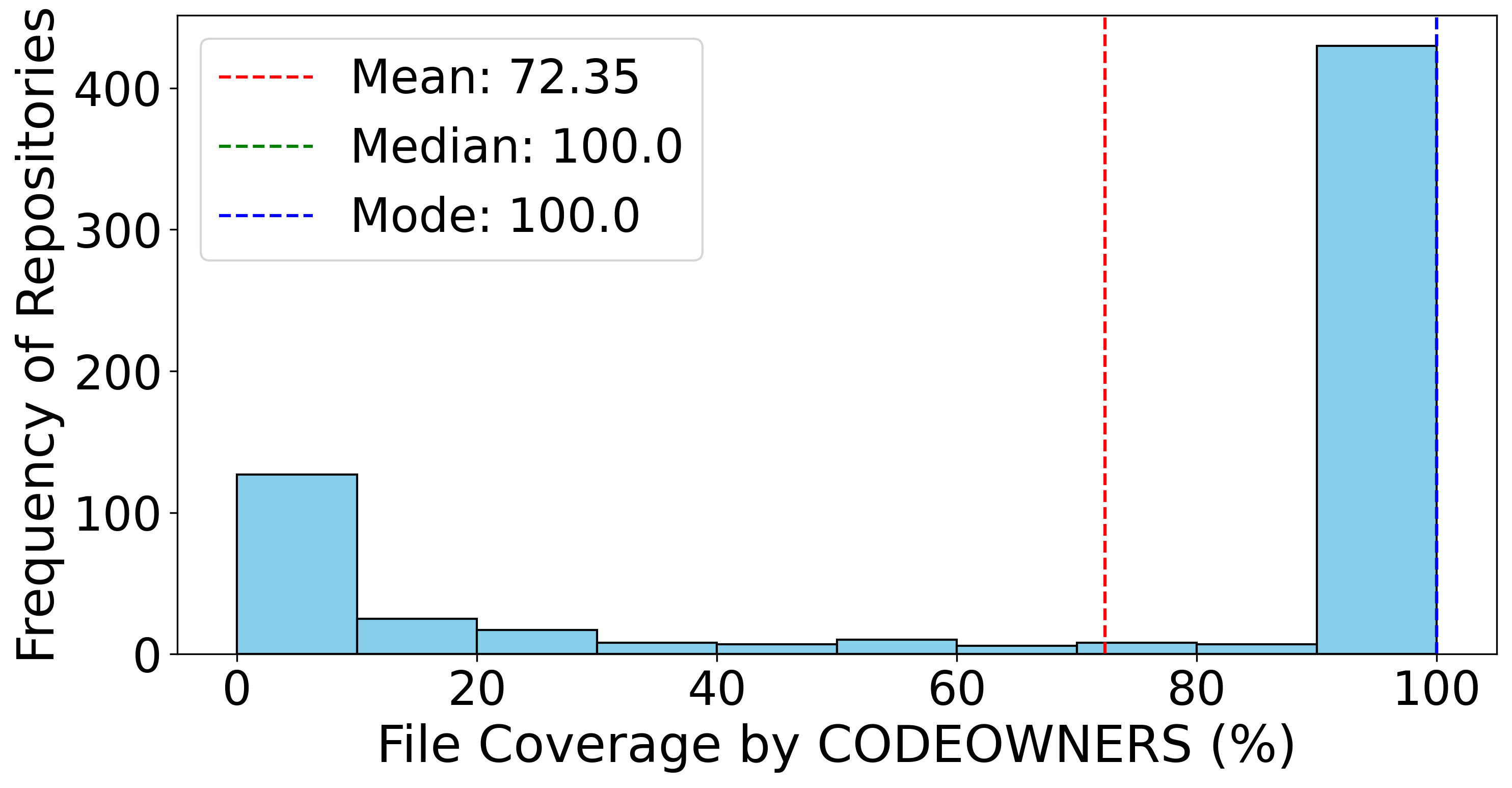}
\caption{Distribution of File Coverage by CODEOWNERS Across Repositories.}
\label{fig:file_coverage_distribution}
\end{subfigure}
\caption{Characteristics of CODEOWNERS files across repositories: rule line counts and file coverage.}
\label{fig:char-of-codeowner-files-across-repos}
\end{figure}

We next analyzed the structure of CODEOWNERS files and the extent of their coverage across repositories. Using the CODEOWNERS Parser tool, we identified the number of ownership rules per repository and calculated the percentage of files covered by these rules.
Most CODEOWNERS files were compact: the median contained only two rules, with an average of 40.45 rules per file (see Figure~\ref{fig:rule_line_counts_distribution}). Remarkably, 40\% of repositories had only a single rule.

As shown in Figure~\ref{fig:file_coverage_distribution}, coverage exhibited a bimodal distribution—repositories typically protected either very little ($\leq$10\% of files) or almost everything ($\geq$90\%). This polarization suggests that projects either use CODEOWNERS to secure a few critical components or adopt it comprehensively as a governance mechanism. \revone[3.I]{By ``critical'' (or key) components we do not mean an arbitrary subset: as Table~\ref{table:patterns_by_count} shows, the protected files are overwhelmingly recurring high-impact, non-code assets including build and CI/workflow configurations, dependency manifests, and licensing or documentation files, whose modification most directly affects project integrity.}

Table~\ref{table:patterns_by_count} lists the most common ownership patterns, which frequently targeted configuration, documentation, and workflow files (e.g., \texttt{.github/}, \texttt{docs/}, \texttt{.github/workflows/}, \texttt{LICENSE}, and \texttt{package.json}). These findings indicate that CODEOWNERS is primarily used to protect non-code assets essential to project infrastructure, such as continuous integration configurations, documentation, and licensing files.

\begin{table}
\centering
\caption{Distribution of Patterns by Count. \revthree[Tbl2]{The \emph{Count} column reports the number of CODEOWNERS files in which each pattern appears at least once, with percentages taken over the 645 files analyzed. The \emph{unique occurrence count} is the number of files in which the pattern is file's only rule (e.g., \texttt{*} is the sole rule in 244 of the 410 files where it appears).}}
\label{table:patterns_by_count}
\begin{tabular}{p{0.28\columnwidth} r r r}
\toprule
\textbf{Pattern} & \textbf{Count} & \textbf{Percentage} & \makecell{\textbf{Unique} \\ \textbf{Occurrence Count}} \\
\midrule

* & 410 & 63.57\% & 244 \\
/.github/ & 48 & 7.44\% & 2 \\
/docs/ & 36 & 5.58\% & 2 \\
*.md & 17 & 2.64\% & 1 \\
.github/ & 17 & 2.64\% & 4 \\
/.github/CODEOWNERS & 16 & 2.48\% & 0 \\
/.github/workflows/ & 14 & 2.17\% & 4 \\
/tools/ & 12 & 1.86\% & 0 \\
/package.json & 10 & 1.55\% & 0 \\
/*.md & 9 & 1.40\% & 0 \\
/scripts/ & 9 & 1.40\% & 0 \\
/LICENSE & 9 & 1.40\% & 0 \\
.gitignore & 9 & 1.40\% & 0 \\
\bottomrule
\end{tabular}
\end{table}

\section{Empirical Study}\label{sec:findings}

Section~\ref{sec:intro} states our four research questions. RQ1 and RQ2 explore adoption and adherence—whether projects follow their ownership rules, and whether declared owners align with contribution-based ownership. RQ3 and RQ4 assess behavioral impact over time, quantifying how introducing CODEOWNERS changes PR processing and workload distribution. Each subsection below combines the methodological detail and the findings specific to its RQ. Because RQ3 and RQ4 both analyze temporal effects before and after adoption, we first introduce the trend analysis technique they share.

\subsection{Trend Analysis Technique}

To assess how the adoption of CODEOWNERS influences repository activity (RQ3, RQ4), we employed a Regression Discontinuity Design (RDD). RDD provides a quasi-experimental framework for estimating causal effects when a clear intervention point—in our case, the introduction of CODEOWNERS—divides the timeline into pre- and post-treatment segments.

We use RDD because repository activity is influenced by many confounding factors—such as contributor turnover, release cycles, and project growth—that make it difficult to isolate the causal impact of CODEOWNERS adoption. \revtwo[S1]{We prefer RDD over the two most common alternatives for this setting. \rtwothree[1]{A difference-in-differences (DiD) design would require a comparison group of repositories that do \emph{not} adopt CODEOWNERS, or have not yet adopted it, and whose pull-request activity follows parallel pre-adoption trends. Staggered adoption across calendar time does not by itself rule out DiD: estimators designed for variation in treatment timing~\cite{callaway2021did,goodmanbacon2021did,roth2023trending} handle it, and He et al.~\cite{he2026cursor} apply such a design to tool adoption in open-source repositories. What our data lacks is the comparison group. We did not mine pull requests for the non-adopting baseline repositories (Section~2.3), and not-yet-adopting repositories would serve as controls only under the untestable assumption that the timing of adoption is unrelated to the trends we measure.}
 Propensity-score matching corrects for selection on \emph{observable} repository characteristics but does not exploit the timing of adoption and still depends on a matched comparison group. RDD instead lets each repository serve as its own control, identifying the effect from the discontinuity at that repository's own adoption point, which directly matches our quasi-experimental question and avoids cross-repository matching assumptions.}
By focusing on local changes around the adoption event, RDD helps approximate causal inference without requiring random assignment. To improve internal validity in the causal estimate, we restricted the analysis window to a balanced 24-month observation period (12 months before and 12 months after CODEOWNERS adoption). \revone[3.IV]{We chose a 12-month half-window to balance bias against variance: a window long enough to capture sustained, seasonally-robust trends rather than transient reactions to adoption, yet short enough that (i) most repositories in our sample retain sufficient PR history on both sides of the cutoff to fit stable local regressions, and (ii) the linear approximation within each segment remains reasonable near the discontinuity. Substantially longer windows would have admitted more confounding from unrelated long-run project changes.} In addition, we excluded the 1-month buffer on both sides of the cutoff to reduce contamination caused by anticipatory behaviors (i.e., teams preparing for CODEOWNERS onboarding) and post-adoption habituation. \rtwoone[1]{Specifically, for each repository and metric we fit a single ordinary-least-squares model in which the running variable (months since adoption) interacts with a treatment indicator, so that the pre- and post-adoption segments have their own intercept and slope; Equations~\eqref{eq_bfr_cutoff} and~\eqref{eq_aft_cutoff} write out the two segments of this one model. The outcome metrics are PR merge time, number of comments, and number of reviewers, each log-transformed.}

We decompose each repository's activity pattern into three interpretable components: pre-slope, cutoff, and post-slope, as this structure is intrinsic to RDD interpretation:
\rtwoone[1]{the pre-slope reflects the underlying trajectory before intervention, the cutoff represents the discontinuity or immediate jump at the intervention point, and the post-slope term measures how the trajectory changes at the intervention relative to the pre-adoption trend.}
This decomposition allows us to determine whether observed changes are abrupt, gradual, or continuations of existing trends.

We controlled for potential confounders that might independently influence PR dynamics, including repository age, size, number of contributors, number of files modified per PR, and overall commenting and reviewing activity. Temporal variables (e.g., time since adoption) were computed dynamically to maintain per-PR accuracy.

\begin{table*}
\centering
\caption{Categorization Process for RDD Analysis}
\label{table:categorization_process_for_RDD_analysis}
\begin{tabular}{l p{0.4\columnwidth} p{0.4\columnwidth} p{0.65\columnwidth} p{0.25\columnwidth}}
\toprule
\textbf{Category} & \textbf{Definition} & \textbf{How It's Determined} & \textbf{Possible Values} & \textbf{Measure Type} \\

\midrule
Pre-Slope 
& The trend before CODEOWNERS adoption, capturing the direction of change
& Evaluated using the slope before cutoff ($\beta_{pre}$) 
&
- Accelerate ($\beta_{pre} >$ Threshold) \newline
- Decelerate ($\beta_{pre} < -$Threshold) \newline
- Constant ($-$Threshold $\le \beta_{pre} \le$ Threshold)
&
Absolute
\\

\midrule
Cutoff 
& The immediate change (jump/drop) at the moment of CODEOWNERS adoption
& Evaluated using the discontinuity at cutoff ($\beta_{cutoff}$) 
&
- Rise ($\beta_{cutoff} >$ Threshold) \newline
- Drop ($\beta_{cutoff} < -$Threshold) \newline
- Flat ($-$Threshold $\le \beta_{cutoff} \le$ Threshold)
&
Relative to intercept ($\beta_{0}$)
\\

\midrule
Post-Slope
& \rtwoone[1]{The change in slope at adoption: the post-adoption slope minus the pre-adoption slope}
& \rtwoone[1]{Evaluated using the interaction coefficient ($\beta_{post}=\beta_{3}$)}
&
\rtwoone[1]{- Accelerate ($\beta_{post} >$ Threshold): trend bends upward relative to before \newline
- Decelerate ($\beta_{post} < -$Threshold): trend bends downward relative to before \newline
- Constant ($-$Threshold $\le \beta_{post} \le$ Threshold)}
&
Relative to $\beta_{pre}$
\\

\bottomrule
\end{tabular}
\end{table*}

Formally, the two fitted models are defined as follows.

\noindent \text{Before cutoff (treatment = 0):} \quad
\begin{equation} \label{eq_bfr_cutoff}
\begin{split}
\hat{y}_{\text{before}} &= \beta_0 + \beta_1 \cdot x_{\text{before}} + \beta_4 \cdot \text{repo\_age} + \beta_5 \cdot \text{repo\_size} \\
&\quad + \beta_6 \cdot \text{num\_comments} \\
&\quad + \beta_7 \cdot \text{num\_review\_comments} \\
&\quad + \beta_9 \cdot \text{num\_reviewers} \\
&\quad + \beta_{10} \cdot \text{files\_modified} \\
&\quad + \beta_{11} \cdot \text{unique\_authors\_count} \\ 
&\quad + \beta_{12} \cdot \text{unique\_mergers\_count} \\
\end{split}
\end{equation}

\noindent \text{After cutoff (treatment = 1):} \quad
\noindent \begin{equation} \label{eq_aft_cutoff}
\begin{split}
\hat{y}_{\text{after}} &= (\beta_0 + \beta_2) + (\beta_1 + \beta_3) \cdot x_{\text{after}} \\
&\quad + \beta_4 \cdot \text{repo\_age} + \beta_5 \cdot \text{repo\_size} \\
&\quad + \beta_6 \cdot \text{num\_comments} \\
&\quad + \beta_7 \cdot \text{num\_review\_comments} \\
&\quad + \beta_9 \cdot \text{num\_reviewers} \\
&\quad + \beta_{10} \cdot \text{files\_modified} \\
&\quad + \beta_{11} \cdot \text{unique\_authors\_count} \\
&\quad + \beta_{12} \cdot \text{unique\_mergers\_count} \\
\end{split}
\end{equation}

\rtwoone[1]{Equations~\eqref{eq_bfr_cutoff} and~\eqref{eq_aft_cutoff} are written for the merge-time outcome. Each outcome is excluded from its own predictors: in the model for the number of comments, $\beta_{6}$ multiplies the number of reviewers and $\beta_{9}$ merge time, and in the model for the number of reviewers, $\beta_{6}$ multiplies the number of comments and $\beta_{9}$ merge time, as in the released implementation.}

\noindent From these models, we derive three key parameters:
\begin{itemize}
    \item $\beta_{pre} = \beta_{1}$
    \item $\beta_{cutoff} = \beta_{2}$
    \item $\beta_{post} = \beta_{3}$
\end{itemize}

These parameters respectively capture the trend before adoption (pre-slope), the immediate jump or drop at adoption (cutoff), and \rtwoone[1]{the change in trend at adoption (post-slope)}.
The categories and interpretations used for these parameters are summarized in Table~\ref{table:categorization_process_for_RDD_analysis}.
\rtwoone[1]{Two slope quantities can be read from Equation~\eqref{eq_aft_cutoff}. The absolute slope of the post-adoption segment is $\beta_{1}+\beta_{3}$; the interaction coefficient $\beta_{3}$ is the change in slope at adoption, that is, the post-adoption slope minus the pre-adoption slope $\beta_{1}$. We classify $\beta_{3}$, so $\beta_{post}=\beta_{3}$ throughout and the post-slope categories in Table~\ref{table:categorization_process_for_RDD_analysis} are relative to the pre-adoption trend: a Decelerate label states that the trend flattened or turned downward at adoption, not that the metric falls afterwards, since a metric that was rising may continue to rise more slowly. We use $\beta_{3}$ because the discontinuities at the cutoff in level ($\beta_{2}$) and in slope ($\beta_{3}$) are the quantities an RDD attributes to the intervention, whereas the absolute slope also carries the pre-existing trend. RQ4 reads the same three quantities from its group-level fits.}

The threshold in Table~\ref{table:categorization_process_for_RDD_analysis} is calculated using \emph{Median Absolute Deviation (MAD)}, which measures the median absolute distance of each data point from the median. MAD was chosen because it is robust and interpretable, as summarized:

\begin{itemize}
    \item It maintains the original scale of the data and is less sensitive to extreme values.
    \item It is robust to skewed distributions and noisy data.
\end{itemize}

To validate this threshold, we manually inspected several repositories with values near and beyond the
computed threshold and observed that those flagged under the 10\%~MAD criterion typically exhibited
meaningful deviations in post-adoption behavior.
\revthree[Tbl6]{The ``Constant'' category is scarce because exact, sustained zero-change around adoption is rare in active repositories; almost every repository exhibits \emph{some} non-zero local slope or jump. Our conclusions depend only on the \emph{sign} of each effect (rise vs.\ drop, accelerate vs.\ decelerate), which is fixed by the estimated coefficient and is insensitive to the precise magnitude of the threshold. The threshold governs only how near-zero a coefficient must be to be labeled ``Constant.'' We confirm this robustness empirically rather than by assertion: Table~\ref{table:threshold_sensitivity} scales the threshold up to $10\times$MAD, which populates the Constant category without acceleration ever overtaking deceleration for merge time and reviews, and restricting the classification to individually significant slopes preserves the same directional conclusion for all three metrics.}
\revtwo[M2]{In plain terms, the categorization reduces to reading the sign of two quantities per repository: \rtwoone[1]{the change in slope at adoption (accelerate / decelerate)} and the size of the jump at adoption (rise / drop), with values close enough to zero called constant.}

\begin{figure}
    \centering
    \includegraphics[width=\columnwidth]{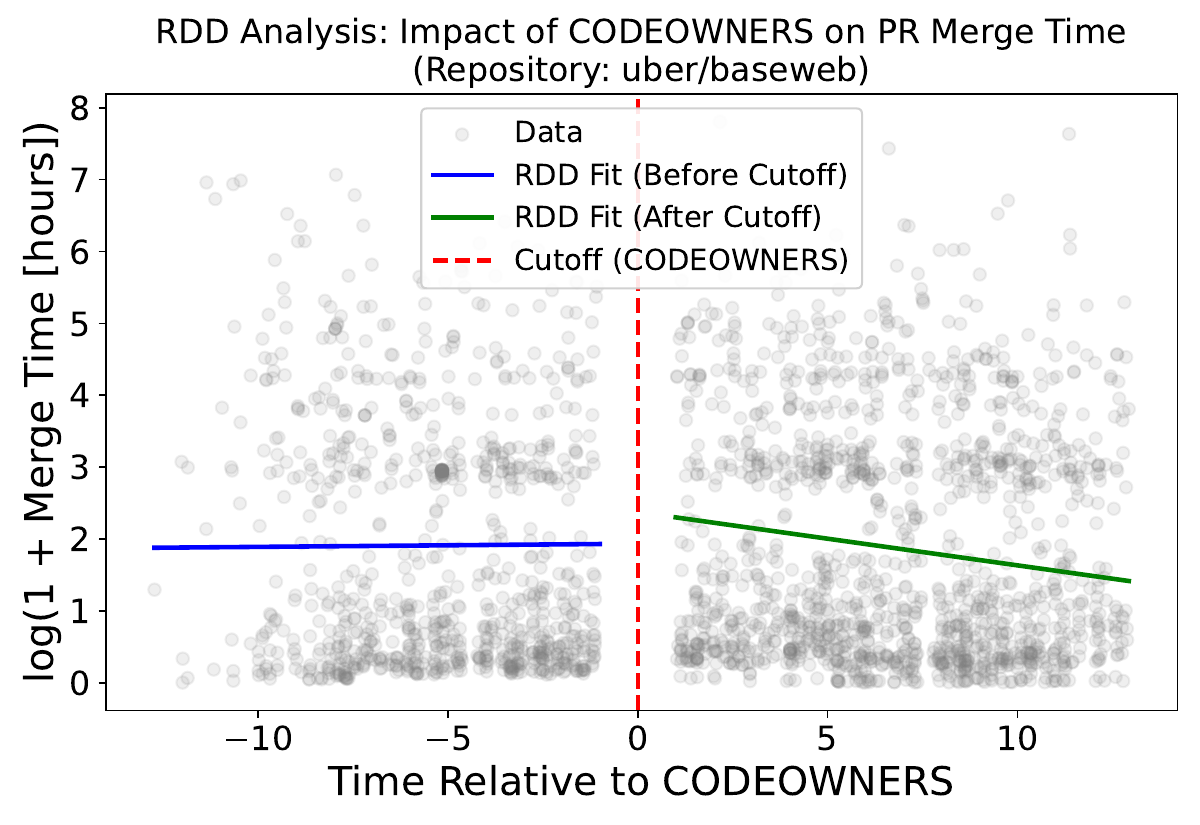}
    \caption{Repository pull request merge time trend before and after CODEOWNERS adoption for \texttt{uber/baseweb}; the outcome is $\log(1+\text{merge time in hours})$. \rtwoone[1]{The fitted segment is approximately flat before adoption ($\beta_{pre}=+0.004$ per month, $p=0.82$) and slopes downward afterwards (absolute post-adoption slope $\beta_{1}+\beta_{3}=-0.074$ per month). The change in slope at adoption, $\beta_{post}=\beta_{3}=-0.079$ per month, is statistically significant ($p<0.001$) and is the quantity the classification uses; the level discontinuity is $\beta_{cutoff}=+0.44$ ($p=0.002$).}}
    \label{fig:rdd_repo_pr_merge_time}
\end{figure}

An example of how categorization is performed is illustrated in
Figure~\ref{fig:rdd_repo_pr_merge_time}, which presents a Regression Discontinuity Design (RDD) plot
showing the impact of CODEOWNERS on pull request (PR) merge time for the repository
\texttt{uber/baseweb}. This repository had 4,342 pull requests, 9,368 comments, 8,455 reviews,
four specified CODEOWNERS, and 150 Git~Blame owners. The blue line represents the fitted regression
line for the period before CODEOWNERS adoption, calculated using
Equation~\ref{eq_bfr_cutoff}, while the green line corresponds to the post-adoption period, calculated
using Equation~\ref{eq_aft_cutoff}. Based on the shape of these regression lines and the definitions in
Table~\ref{table:categorization_process_for_RDD_analysis}, this repository is categorized as:

\begin{itemize}
    \item \textbf{Pre-Slope:} \emph{Accelerate}, inferred from the slope of the blue line ($\beta_{pre}$).
    \item \textbf{Cutoff:} \emph{Rise}, identified from the vertical shift at the cutoff point ($\beta_{cutoff}$).
    \item \textbf{Post-Slope:} \emph{Decelerate}, \rtwoone[1]{inferred from the change in slope between the blue and the green line ($\beta_{post}=\beta_{3}=-0.079$ per month, $p<0.001$): the trend bends downward at adoption.}
\end{itemize}

\revone[3.IX]{Individual pull requests scatter widely across the window, including at its right-hand edge, where merge times in the final month range from minutes to several days. Each fitted line is estimated from all pull requests on its side of the cutoff rather than anchored to any single observation, so the categorization above rests on \rtwoone[1]{the change in slope at the cutoff ($\beta_{3}=-0.079$ per month, $p<0.001$)} and not on individual points.}

\begin{figure*}
    \centering
    \includegraphics[width=1.33\columnwidth]{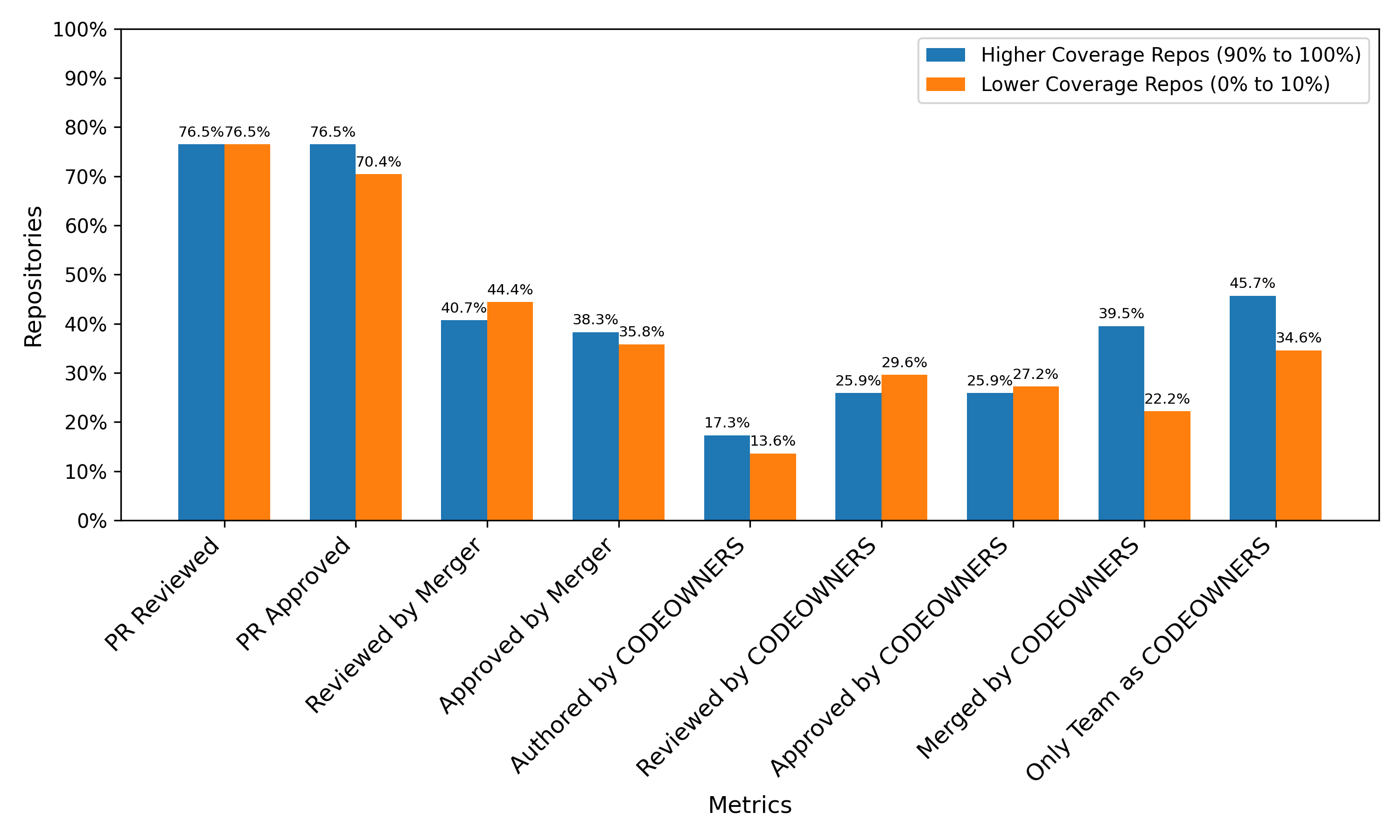}
    \caption{Code owner participation at each pull request stage in repositories with higher coverage ($\geq$90\%) versus lower coverage ($\leq$10\%). \revtwo[S3]{Each bar is the percentage of the 81 repositories in the group whose sampled pull request had the property shown.}}
    \label{fig:codeowners_activties_comparison} 
\end{figure*}

\subsection{Results}

\subsubsection{RQ1 — Adherence to CODEOWNERS}

\begin{table}
\centering
\caption{Median repository characteristics of low-coverage ($\leq$10\%, $n=127$) and high-coverage ($\geq$90\%, $n=430$) CODEOWNERS repositories, with two-sided Mann--Whitney U tests and Cliff's $\delta$ effect sizes.}
\label{table:coverage_group_characteristics}
\begin{tabular}{lrrcc}
\toprule
\textbf{Metric} & \textbf{Low} & \textbf{High} & \textbf{$p$} & \textbf{Cliff's $\delta$} \\
\midrule
Contributors  & 307   & 177   & $<0.001$ & 0.23 \\
Commits       & 8,339 & 3,156 & $<0.001$ & 0.33 \\
Pull requests & 4,313 & 1,627 & $<0.001$ & 0.28 \\
Issues        & 2,585 & 1,123 & $<0.001$ & 0.23 \\
Age (years)   & 9.3   & 8.3   & $0.009$  & 0.15 \\
\bottomrule
\end{tabular}
\end{table}

To examine how consistently CODEOWNERS rules are followed in practice, we analyzed repositories with valid CODEOWNERS files.
As established in our preliminary findings, repositories tend to cluster into two distinct categories based on file coverage: Low Coverage (0-10\%) and High Coverage (90-100\%). To ensure a balanced comparison, we randomly selected an equal number of repositories from each group (n = 81 per group) and extracted pull request (PR)-level data to evaluate how often designated owners were involved in key development activities.
\revtwo[S3]{\rtwoone[3]{From each sampled repository we took the most recent merged PR that postdated the most recent modification of the CODEOWNERS file and modified at least one owned file, and we recorded which ownership actions were performed on it. We stopped at one PR per repository because each candidate costs several API requests under GitHub's rate limit.} Each repository therefore contributes one observation per stage, and the percentages reported for these two samples are shares of repositories rather than shares of PRs. The full-dataset rates given at the end of this subsection are the exception: those are computed over all PRs in all analyzed repositories.}
\revtwo[S3]{This two-group design reflects the bimodal coverage distribution shown in Figure~\ref{fig:file_coverage_distribution}: of the 645 repositories with CODEOWNERS files, 127 fall in the low-coverage band and 430 in the high-coverage band, leaving 88 in the intermediate 10--90\% range. We compare the two dominant modes and do not separately analyze this intermediate group, which, though smaller, may reflect more deliberate adoption where owners specify targeted rules rather than a single blanket rule. Whether adherence increases with coverage across this middle range is a question we leave to future work.}

\revtwo[M3]{The two bands also differ as projects, not only in coverage level. As Table~\ref{table:coverage_group_characteristics} shows, across the 127 low- and 430 high-coverage repositories the low-coverage group is consistently larger and marginally older, with more contributors, commits, pull requests, and issues. All five differences are statistically significant, with modest effect sizes (Cliff's $\delta$ between 0.15 and 0.33). This fits the two adoption styles noted earlier: the largest and most active projects tend to protect only a few critical files, producing low coverage, whereas smaller projects more often place the whole repository under a single blanket rule, producing high coverage.}


We defined a set of metrics to quantify adherence across multiple participation types:
\begin{itemize}
\item \textit{Reviewed / Approved}: Whether the PR received any review or approval from any contributor.
\item \textit{Reviewed / Approved / Merged / Authored by CODEOWNER}: Whether a designated code owner participated in these actions.
\item \textit{Reviewed / Approved by Merger}: Whether the individual who merged the PR also reviewed or approved.
\item \textit{Team-only Ownership}: Whether all affected files were owned exclusively by teams rather than individuals.
\end{itemize}

Using these metrics, we define \textbf{adherence} as follows: a PR adheres to CODEOWNERS if at least one \rtwoone[2]{ownership action} (review, approval, or merge) is performed by a designated code owner. We also consider a relaxed variant of this definition in which authorship by a code owner is additionally counted as an ownership action.
\rtwoone[2]{We use the term adherence for brevity: the measure records whether a designated owner took part in the PR, not whether the repository required it through a branch protection rule or ruleset (Section~\ref{sec:threats}).}

These indicators capture both direct adherence to ownership policies—cases where CODEOWNERS perform the expected review, approve or merge actions—and indirect adherence, where responsibilities are fulfilled by others.

Figure \ref{fig:codeowners_activties_comparison} compares code owner participation across the two coverage groups.
In the low-coverage sample ($\leq$10\% of files protected), participation by code owners in specific actions was modest.
A code owner authored the sampled PR in 14\% of these repositories,
reviewed it in 30\%,
approved it in 27\%, and
merged it in 22\%.
These values suggest that when CODEOWNERS coverage is minimal, projects rely on informal review processes or broader contributor participation.
In contrast, high-coverage repositories ($\geq$90\% of files protected) showed higher owner involvement at the merge stage, with comparable review and approval rates:
an owner authored the sampled PR in 17\% of these repositories,
reviewed it in 26\%,
approved it in 26\%, and
merged it in 40\%.
\revtwo[S3]{The main difference between the groups is at the merge stage, where an owner merged the sampled PR in 40\% of high-coverage repositories versus 22\% of low-coverage ones; authoring is modestly higher (17\% versus 14\%) and review and approval are broadly similar. Treating each repository's owner involvement at a stage as a binary outcome, we test these differences with Fisher's exact test on the size-matched samples (81 repositories per group; Appendix~\ref{app:adherence_tests}). Only the merge-stage difference is statistically significant \rtwoone[3]{in this sample} (\emph{p}~=~0.027, odds ratio~=~2.29, Cohen's \emph{h}~=~0.38); the authoring, review, and approval differences are not (\emph{p}~$\geq$~0.66). \rtwoone[3]{In this sample,} broader CODEOWNERS coverage is thus associated with more consistent owner \rtwoone[2]{involvement} specifically at the final integration (merge) stage\rtwoone[3]{; the additional analysis over all eligible pull requests (Section~\ref{sec:threats}) does not provide strong support for this difference}.}

\revone[3.III]{These raw differences, however, are largely a structural effect of coverage rather than a behavioral one. Coverage strongly determines how often a pull request even touches an owned file: only 11\% of pull requests in low-coverage repositories modify an owned file, versus 93\% in high-coverage repositories. Considering all pull requests, a code owner merged 21\% of them in high-coverage repositories but only 2\% in low-coverage ones (Cohen's $h=0.68$, a large effect). Conditional on a pull request actually modifying an owned file, that is, when an owner is genuinely responsible, the difference collapses to a small effect: owners merged 19\% of such pull requests under low coverage and 23\% under high coverage ($h=0.11$), and the review, approval, and authoring differences are likewise negligible ($h\leq0.13$; Appendix~\ref{app:coverage_control}). Broader coverage therefore raises owner involvement chiefly by placing owners on more pull requests, as the mechanism intends, rather than by making owners more diligent on the pull requests they already own.}

When we aggregate these action-level observations into our adherence definition, we find that high-coverage repositories exhibit somewhat stronger but still far-from-perfect adherence. Under the strict definition (only reviews, approvals, and merges by code owners count as \rtwoone[2]{ownership actions}), the sampled PR adheres to CODEOWNERS in 41.98\% of high- and 38.27\% of low-coverage repositories.

Under the relaxed definition, where authorship by a code owner also counts as an ownership action, adherence in high-coverage repositories increases to 46.91\% of repositories, while adherence in low-coverage repositories remains at 38.27\%. This stability in the low-coverage group indicates that, in these projects, code owners rarely contribute \emph{only} as authors in cases where they are not already involved in review, approval, or merge; authorship thus adds little additional adherence signal there. Measured across all PRs in the full dataset rather than across the sampled repositories, the strict definition yields an overall adherence rate of 19.07\% and the relaxed definition 20.31\%.

These results suggest that\rtwoone[3]{, in the sampled repositories,} broader CODEOWNERS coverage is associated with more systematic involvement of designated owners, especially at the merge stage, and with moderately higher adherence\rtwoone[3]{, although the additional analysis over all eligible pull requests (Section~\ref{sec:threats}) does not provide strong support for these differences}. \rtwoone[2]{However, in more than half of the high-coverage repositories the sampled PR does not satisfy our adherence criteria.}

\summarybox{RQ1}{Adherence to CODEOWNERS rules in repositories}{
\rtwoone[2]{Designated owners take part in the pull requests that touch their files in a non-negligible but far-from-universal way.} Among the sampled repositories with $\leq$10\% coverage, a code owner authored the sampled PR in 14\%, reviewed it in 30\%, approved it in 27\%, and merged it in 22\%. Among those with $\geq$90\% coverage, the corresponding shares are 17\%, 26\%, 26\%, and 40\%. Adherence, defined as at least one \rtwoone[2]{ownership action} (review, approval, or merge) by a code owner, holds in 41.98\% of high-coverage versus 38.27\% of low-coverage repositories; measured over all PRs in the full dataset, the overall adherence rate is 19.07\%. When authorship is additionally counted as an ownership action, adherence in high-coverage repositories rises to 46.91\% (with low-coverage remaining at 38.27\%), and the overall PR-level rate increases slightly to 20.31\%. \rtwoone[2]{Broader coverage thus corresponds to more frequent, though still partial, owner involvement at the merge stage.}}


\begin{figure}
\centering
\begin{subfigure}[b]{0.5\textwidth}
\centering
\includegraphics[width=0.9\columnwidth]{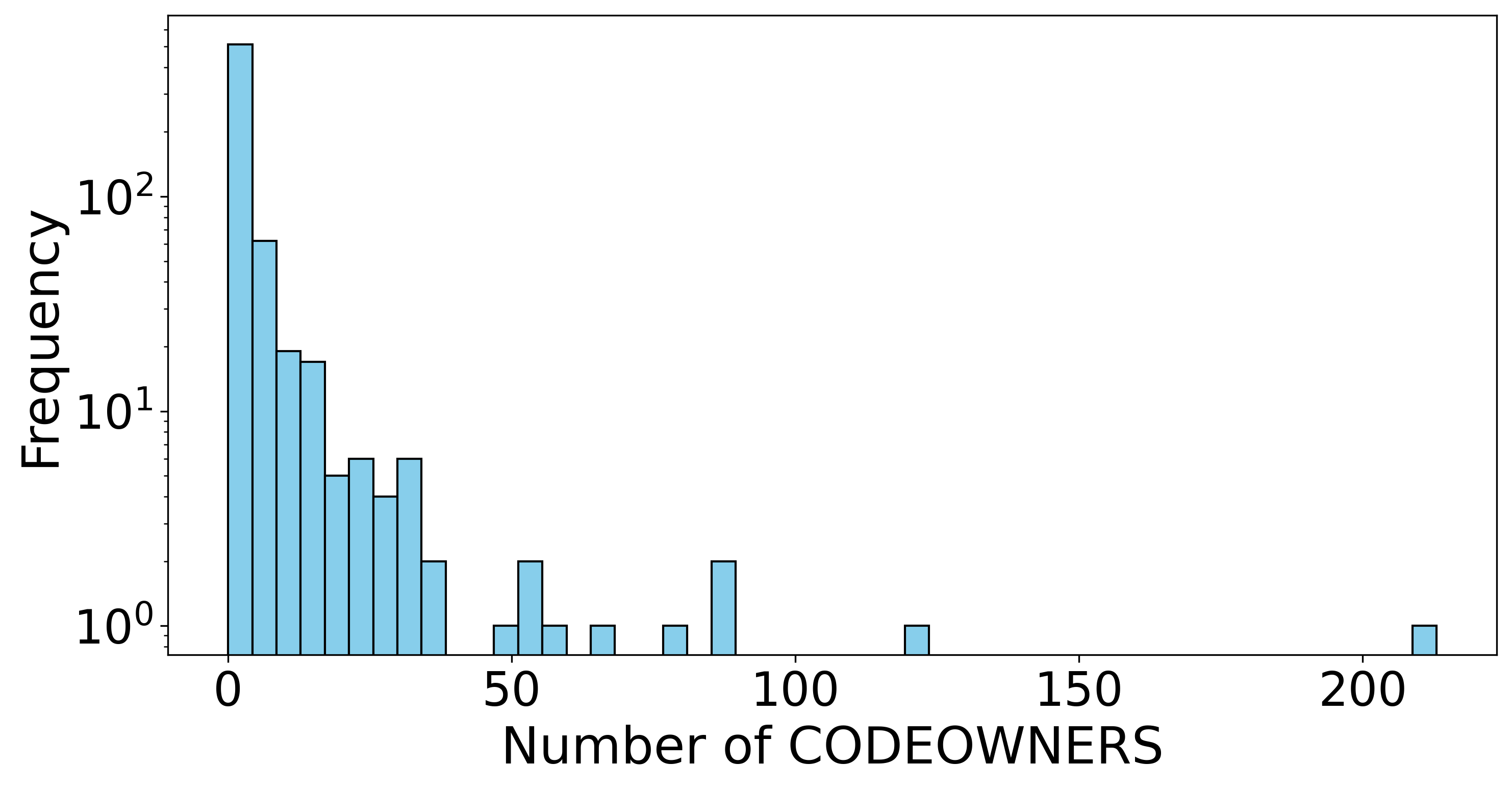}
\caption{Unique code owners per Repository (three outliers omitted).}
\label{fig:unique_codeowners_distribution}
\end{subfigure}
\begin{subfigure}[b]{0.5\textwidth}
\centering
\includegraphics[width=0.85\columnwidth]{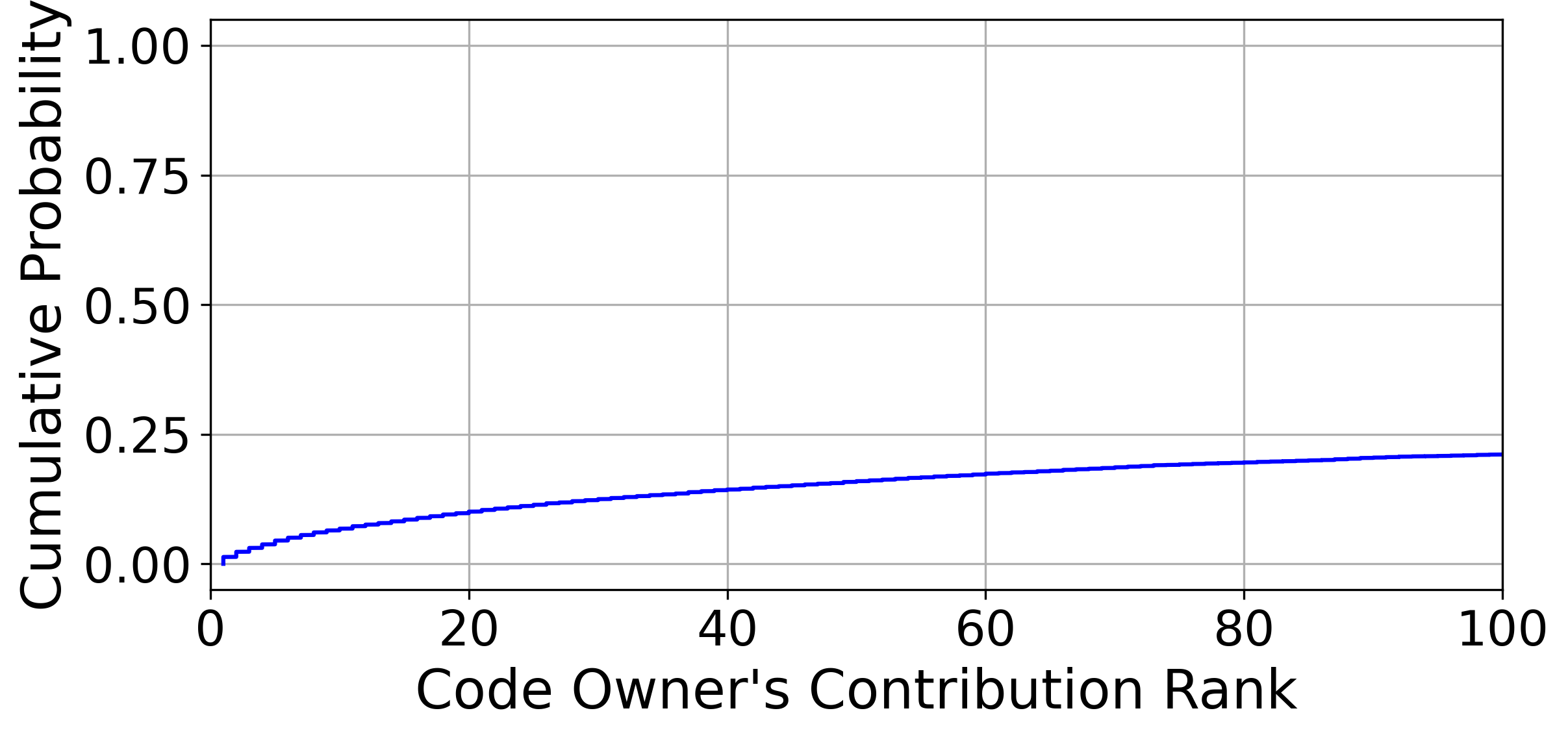}
\caption{Cumulative distribution of individual code owners by contribution rank (rank 1 = most frequent committer). The curve reaches only 0.21 at rank 100 because the remaining 79\% of code owners fall outside the top 100 committers and so have no rank in this range.}
\label{fig:codeowner_rankings_cdf}
\end{subfigure}
\caption{Characteristics of the code owners declared across repositories: how many distinct owners each repository names, and where those owners fall in their repository's contribution ranking.}
\label{fig:char-of-codeowners-across-repos}
\end{figure}

\begin{table*}
\centering
\caption{Repository CODEOWNERS Activities Not in the Top 100 Contributors (2022–2024).}
\label{table:codeowner_sample}
\begin{tabular}{llccc}
\toprule
\textbf{Repository} & \textbf{Member} & \textbf{Commits} & \textbf{Opened PRs} & \textbf{Reviewed PRs} \\
\midrule
\texttt{tensorflow/tensorflow} & qqfish & 0 & 0 & 4 \\
\texttt{tensorflow/tensorflow} & azaks2 & 0 & 0 & 0 \\
\texttt{ohmyzsh/ohmyzsh} & septs & 2 & 1 & 4 \\
\texttt{home-assistant/core} & devbis & 2 & 2 & 0 \\
\texttt{DefinitelyTyped/DefinitelyTyped} & aarystan & 0 & 0 & 0 \\
\texttt{google/comprehensive-rust} & voss & 0 & 0 & 1 \\
\texttt{bazelbuild/bazel} & pzembrod & 40 & 1 & 18 \\
\texttt{redis/go-redis} & dmaier-redislabs & 0 & 0 & 1 \\
\texttt{NixOS/nixpkgs} & lukegb & 340 & 46 & 127 \\
\texttt{mono/mono} & SamMonoRT & 0 & 0 & 0 \\
\bottomrule
\end{tabular}
\end{table*}


\subsubsection{RQ2 — Comparison Against Contribution-Based Ownership Metrics}

Contribution-based code ownership in software engineering is typically inferred from developers' contribution histories rather than declared through explicit policy.
To contrast such implicit ownership with GitHub's explicit CODEOWNERS mechanism, we first examine \emph{who} the declared owners are, and then compare \emph{how} they behave in pull requests relative to contribution-based owners.

We began by characterizing the population of declared CODEOWNERS in our dataset.
For each repository, we retrieved the top 100 contributors via the GitHub API and determined whether individuals listed in CODEOWNERS files were among them.
\revone[3.V]{We distinguish individual from team owners directly from CODEOWNERS syntax: entries of the form \texttt{@org/team} denote a GitHub team, whereas \texttt{@username} entries denote an individual account. The owner counts reported here and in Figure~\ref{fig:unique_codeowners_distribution} are derived from this parsed classification, and only individual accounts are matched against the top-100 contributor list (teams are excluded from the ranking comparison because their membership is not publicly visible).}
As illustrated in Figure~\ref{fig:unique_codeowners_distribution}, the median repository listed one individual code owner (mean 15.95) and few team owners (mean 1.8).
More strikingly, \revtwo[M4]{Figure~\ref{fig:codeowner_rankings_cdf} plots the cumulative distribution of individual code owners over their contribution rank; reading specific points from this curve,} 79\% of individual code owners (8,114 out of 10,287) were not among the top 100 most frequent committers, and only 1\% (137 individuals) were the most frequent committer in their repositories. 
Manual inspection of randomly sampled cases (Table~\ref{table:codeowner_sample}) further suggested that many such owners are more active in reviewing than coding.
Taken together, these observations indicate that CODEOWNERS often represent a governance-oriented layer—reviewers, maintainers, or organizational representatives—rather than the core group of contributors who make most commits.

To construct a comparable contribution-based notion of ownership, we define a contribution-based owner as a developer who has contributed at least 5\% of a file's total lines of code (LOC), following Foucault et al.~\cite{foucalEtAll2014}.
This threshold captures sustained responsibility through ongoing modification and maintenance, consistent with prior empirical studies of implicit ownership in large-scale codebases.
Using Git's \texttt{blame} functionality, which attributes each line of code to a specific author, we identified such contribution-based owners and mapped the extracted author information—typically email-based—to GitHub usernames on a best-effort basis.
Some mappings remained incomplete due to GitHub's privacy settings (e.g., “hide email” option), but from the successfully mapped data we obtained contribution-based owners at the file, pull request (PR), and repository levels.

For each PR, we then determined the roles played by both groups of owners—those explicitly listed in CODEOWNERS files and those inferred from \texttt{git blame}.
Using a consistent classification procedure, we identified whether each individual acted as an author, reviewer, approver, or merger.
We compared their behaviors using the Mann-Whitney U test to assess distributional differences, and Cliff's $\delta$ to estimate the practical effect size.
This analysis directly tests whether ownership defined by policy (CODEOWNERS) differs in practice from ownership inferred by contribution history, focusing both on the distinct identities of these groups and on whether they behave differently in review and merge activities.

\begin{table}
\centering
\caption{Behavioral comparison of CODEOWNERS against contribution-based owners
on four per-pull-request collaboration metrics. Every effect size is negligible
by the $\delta<0.147$ threshold of Romano et al.~\cite{romano:2006}.}
\label{table:codeowners_vs_blame_owners}
\begin{tabular}{lccc}
\toprule
\textbf{Metric} & \makecell{\textbf{Mann--Whitney}\\\textbf{\emph{U}} ($\times10^{9}$)}
& \textbf{\emph{p-value}} & \textbf{Cliff's $\delta$} \\
\midrule
Comments        & 14.53 & $<0.001$ & 0.020 \\
Review comments & 14.67 & $<0.001$ & 0.010 \\
Reviews         & 13.48 & $<0.001$ & 0.090 \\
Merge time      &  9.74 & $<0.001$ & 0.034 \\
\bottomrule
\end{tabular}
\end{table}

While all Mann-Whitney U tests revealed statistically significant differences ($p<0.001$), all Cliff's $\delta$ values were negligible (Table~\ref{table:codeowners_vs_blame_owners}), suggesting that the practical differences between the groups' behaviors are minimal.
In other words, even though CODEOWNERS and contribution-based owners are often \emph{different people}—with CODEOWNERS frequently drawn from outside the core contributor set—their observable collaborative behaviors in PRs (commenting, reviewing, and merging) are very similar.
\revone[3.VII]{This similarity is \emph{conditional on participation}: the comparison is computed over PRs in which each type of owner is actually involved, so it describes how owners behave \emph{when} they engage, not how often. Code owners are usually not the most frequent PR participants and engage less often than core contributors, yet the per-PR footprint of their participation is comparable to that of contribution-based owners.}

Overall, these results show a clear separation between identity and behavior.
CODEOWNERS assignments extend ownership beyond core committers and toward governance roles, yet when these declared owners participate in PRs, they act much like contribution-based owners in terms of engagement and decision-making.
Rather than creating entirely new behavioral patterns, CODEOWNERS formalizes a governance layer whose members, once involved, behave similarly to those who own code through contributions.

\revtwo[S4]{Our comparison captures the \emph{roles} owners play in a pull request: author, reviewer, approver, or merger, and how often they play them, but not the \emph{nature} of the contributions themselves. Two groups can take on similar roles at similar rates yet still differ in what those contributions involve: the type of change (feature, fix, refactor, or configuration), its size and complexity, or the thoroughness of a review. Because our metrics are behavioral rather than content-based, the RQ2 similarity is best read as similarity in observable review and merge participation, not as equivalence in the substance of what owners contribute. Characterizing these qualitative differences would require content analysis of the changes and reviews, which we leave to future work.}

\summarybox{RQ2}{Differences Between CODEOWNERS and Contribution-Based Ownership Metrics}{
Declared CODEOWNERS and contribution-based owners differ substantially in identity—many CODEOWNERS are not among the most frequent committers and often occupy governance-oriented roles—yet they exhibit similar collaborative behaviors in Pull Requests (e.g., reviewing and merging).
Contribution-based owners were defined as contributors responsible for $\geq$5\% of a file's LOC (via \texttt{git blame}).}


\begin{table*}
\centering
\caption{Observed Pull Requests Metric Trends Across Repositories Following CODEOWNERS Adoption. \revthree[Tbl6]{Percentages are computed over all analyzed repositories; within each panel, Rise/Drop/\rtwoone[1]{Flat} (and Accelerate/Decelerate/Constant) entries do not sum to 100\% because repositories whose change was not statistically significant \small{($\alpha=0.05$, $\sim$65\%)} are omitted.}}
\label{table:developers_workload_metric_trend_post_adoption}
\begin{tabular}{lcccccc}
\toprule
\textbf{Metric} & \multicolumn{3}{c}{\textbf{At Cutoff}} & \multicolumn{3}{c}{\textbf{Post Cutoff}} \\
 & \textbf{Rise} & \textbf{Drop} & \textbf{\rtwoone[1]{Flat}} & \textbf{Accelerate} & \textbf{Decelerate} & \textbf{Constant} \\ 
\midrule
\texttt{Merge Time} & 17.2\% & 18.2\% & 0\% & 13\% & 19.3\% & 3.1\% \\  
\texttt{Number of Comments} & 13.3\% & 20.9\% & 0\% & 14.3\% & 17.9\% & 3.1\% \\ 
\texttt{Number of Reviews} & 17.9\% & 15.8\% & 0.5\% & 11.2\% & 20.9\% & 2.1\% \\ 
\bottomrule
\end{tabular}
\end{table*}

\vspace{-2em}
\subsubsection{RQ3 — Impact of CODEOWNERS Adoption on Pull Request Dynamics}

To assess how the introduction of CODEOWNERS \rtwoone[1]{relates to the dynamics of the pull request process}, we analyzed three pull request (PR)-level metrics: merge time, number of comments, and number of reviews.
These indicators capture review effort and turnaround time—two dimensions central to understanding how ownership policies shape the pull request process.
To estimate changes attributable to CODEOWNERS adoption, we employed a Regression Discontinuity Design (RDD), as detailed in Equations~\eqref{eq_bfr_cutoff} and~\eqref{eq_aft_cutoff}.
We applied this analysis across 222 repositories that had adopted CODEOWNERS, capturing both immediate (cutoff) effects and longer-term (post-adoption) trends.

Table~\ref{table:developers_workload_metric_trend_post_adoption} summarizes the directional patterns observed in the three metrics.
\rtwoone[1]{Following Table~\ref{table:categorization_process_for_RDD_analysis}, each repository was classified at the cutoff as Rise, Drop, or Flat, and on the post-slope as Accelerate (trend bent upward at adoption relative to the pre-adoption trend), Decelerate (bent downward), or Constant.}

At the adoption cutoff, repositories exhibited mixed short-term effects.
For merge time, 17.2\% of repositories showed an increase, 18.2\% a decrease, and none \rtwoone[1]{were flat}.
For the number of comments, most repositories (20.9\%) experienced a decline, while 13.3\% saw an increase.
Similarly, the number of reviews displayed a balanced distribution (17.9\% rise, 15.8\% drop).
These findings suggest that the immediate aftermath of CODEOWNERS adoption varied by project—some repositories may have initially experienced added coordination overhead as new ownership rules were enforced.

\rtwoone[1]{In the longer term, however, a clearer pattern emerged. More repositories bent downward (Decelerate) than upward (Accelerate) at adoption: merge time in 19.3\% of repositories versus 13\%, comments in 17.9\% versus 14.3\%, and reviews in 20.9\% versus 11.2\%. Decelerate means that a metric which was falling fell faster, or that one which was rising rose more slowly; in either case, after adoption the trends in merge time and review effort bent downward relative to their earlier course. Although roughly 65\% of repositories showed no statistically significant change at the $\alpha = 0.05$ level, the alignment of direction across metrics points to a systematic change in PR dynamics following CODEOWNERS adoption.}
\revthree[6:19]{Because we fit and test one model per repository, the analysis necessarily involves many hypothesis tests. We deliberately do \emph{not} aggregate these into a single family-wise hypothesis or claim that any particular repository's change is significant; instead, our conclusions rest on the \emph{consistency of direction} across the repository population---an argument closer to a sign test over repositories than to a single multiple-comparison family. Per-repository significance is reported only as descriptive context for this directional pattern, not as the basis of our inference, so no family-wise $\alpha$ correction is required for the aggregate claims. The one place we draw an inferential conclusion from a single comparison (Table~\ref{table:codeowners_vs_blame_owners}) involves only four pre-specified metrics, for which a correction would not alter the already-negligible effect sizes. As a further check, we reran the per-repository tests with a Benjamini--Hochberg false-discovery-rate correction ($q = 0.05$). Within each metric the majority direction is unchanged; repositories with greater owner review participation after adoption, for instance, remain the majority. Our aggregate conclusions therefore do not rest on the absence of a correction.}

\revthree[Tbl6]{To confirm that these directional findings do not depend on the specific near-zero threshold, we repeated the classification while scaling the threshold from its default of $0.1\times$MAD up to $10\times$MAD (Table~\ref{table:threshold_sensitivity}). A larger threshold moves more repositories into the Constant category, whose share grows from roughly 2--3\% to about a third; the scarcity of Constant at the default setting therefore reflects the small threshold rather than a forced binary split. The directional signal is stable where it is strongest: for merge time and number of reviews, deceleration outnumbers acceleration at every threshold up to $5\times$MAD, beyond which almost all repositories fall into the Constant category and no direction dominates. For number of comments the two directions are close and their ordering shifts between adjacent thresholds, which marks it as the weakest of the three signals. As a stricter check aimed at the concern that the sign of a near-zero coefficient may itself be unstable, we further restricted the classification to slopes that are statistically significant at $\alpha=0.05$, excluding the near-zero slopes entirely; deceleration remains the majority direction for all three metrics (merge time 23 versus 18 repositories, comments 24 versus 22, reviews 34 versus 6).}


\revtwo[S5]{A natural alternative reading is that \rtwoone[1]{these decelerations} reflect \emph{reduced contribution} rather than \rtwoone[1]{a change in how pull requests are processed}: if adopting CODEOWNERS adds governance overhead that discourages participation, fewer pull requests could shorten merge times without any real gain in throughput. Two observations argue against this. First, the three metrics are measured \emph{per pull request}, so they are rates that the number of pull requests does not mechanically inflate or deflate. Second, absolute activity rose rather than fell after adoption: across the study repositories, the twelve months after adoption contain about a third more pull requests than the twelve months before (185,288 versus 138,593), and at the repository level volume was higher after adoption in 63\% of repositories and lower in 36\% (median ratio 1.17). Excluding bot-authored pull requests leaves the same picture (a 28\% increase; higher in 60\% of repositories). \rtwoone[1]{The deceleration in merge time therefore coincides} with \emph{more} contribution, not less, the opposite of the lowered-throughput scenario. We cannot rule out finer compositional shifts in what is contributed (see the kind-of-contribution limitation for RQ2), and we leave a per-metric decomposition of absolute levels to future work.}

\summarybox{RQ3}{Impact of CODEOWNERS on Pull Request Dynamics}{\rtwoone[1]{Following CODEOWNERS adoption, more repositories decelerated than accelerated, that is, the trend bent downward rather than upward relative to its pre-adoption course, for merge time (19.3\% versus 13\%), comments (17.9\% versus 14.3\%), and reviews (20.9\% versus 11.2\%). While roughly 65\% of changes were not statistically significant, the consistent direction suggests that after adoption the trends in merge time and review effort bend downward relative to their earlier course.}}

\begin{table*}
\centering
\caption{\rtwoone[1]{RDD estimates of pooled owner activity around CODEOWNERS adoption, reported as the three quantities of Table~\ref{table:categorization_process_for_RDD_analysis} (pre-slope $\beta_{pre}$, cutoff $\beta_{cutoff}$, post-slope $\beta_{post}$). Counts are the group's actions per half-month bin over the 13 months on either side of adoption, excluding the month on either side; slopes are in actions per bin per month. The last two columns compare the twelve months after adoption with the twelve months before: the ratio of the group's actions, and that ratio divided by the after/before ratio of pull requests in the repositories where the group is observed, that is, the change in the group's activity per pull request. Pull requests rose by a factor of 1.27 in the 139 repositories whose CODEOWNERS files name individual developers (designated owners) and by 1.34 in all 222 (contribution-based owners). Bold marks cutoff jumps with $p<0.05$; no slope term is statistically significant.}}
\label{table:prs_combined_metrics_trend_post_adoption_codeowners_detailed}
\begin{tabular}{lrrrcc}
\toprule
\textbf{Activity} & \makecell{\textbf{Pre-slope}\\$\beta_{pre}$ ($p$)} & \makecell{\textbf{Cutoff}\\$\beta_{cutoff}$ ($p$)} & \makecell{\textbf{Post-slope}\\$\beta_{post}$ ($p$)} & \makecell{\textbf{Activity}\\(after/before)} & \makecell{\textbf{Activity per PR}\\(after/before)} \\
\midrule
\multicolumn{6}{l}{\textbf{CODEOWNERS} ($n=704$)} \\
Review & $-32.2$ (0.74) & \textbf{+2,686} (0.017) & $+140.9$ (0.32) & 1.61 & 1.26 \\
Decision & $-3.0$ (0.93) & \textbf{+1,343} (0.001) & $-6.7$ (0.90) & 1.73 & 1.36 \\
Authoring & $-8.6$ (0.52) & \textbf{+392} (0.011) & $+0.8$ (0.97) & 1.36 & 1.07 \\
\midrule
\multicolumn{6}{l}{\textbf{Contribution-based owners} ($n=3{,}121$)} \\
Review & $-263.4$ (0.38) & +4,919 (0.15) & $+70.3$ (0.87) & 1.06 & 0.80 \\
Decision & $-81.3$ (0.36) & \textbf{+2,294} (0.022) & $-6.3$ (0.96) & 1.19 & 0.89 \\
Authoring & $-27.2$ (0.60) & \textbf{+1,164} (0.046) & $-39.1$ (0.60) & 1.14 & 0.85 \\
\bottomrule
\end{tabular}
\end{table*}

\vspace{-2.5em}
\subsubsection{RQ4 — Impact of CODEOWNERS on Reviewer Workloads}

To examine how the adoption of CODEOWNERS affects \rtwoone[4]{developer workload}, we analyzed activity patterns before and after adoption across two distinct developer groups: CODEOWNERS (explicitly assigned) and contribution-based Owners (inferred from contribution history).
We focused on three categories of developer engagement that collectively capture workload allocation in the pull request (PR) process:

\begin{itemize}
\item \textbf{Authoring Activity:} Instances where a developer authors a PR.
\item \textbf{Review Activity:} Any engagement in reviewing: comments, formal reviews, or review comments.
\item \textbf{Decision Activity:} Actions involving explicit approval or merging of PRs.
\end{itemize}

Similar to RQ3, we use RDD to assess both immediate changes at the time of CODEOWNERS adoption and longer-term trends in developer activity.
This method accounts for potential confounding factors (as detailed in Section~4.1) \rtwoone[1]{and lets us separate abrupt changes at adoption from gradual changes in trend.}
\rtwoone[1]{Here the unit of analysis is the pooled activity of a developer group rather than a single repository: we count the group's actions in half-month bins relative to adoption, exclude the month on either side of adoption, and fit one model per group and activity category. Table~\ref{table:prs_combined_metrics_trend_post_adoption_codeowners_detailed} reports the three quantities of Section~4.1 for each fit, pre-slope $\beta_{pre}$, cutoff $\beta_{cutoff}$, and post-slope $\beta_{post}$, with $p$-values.}

\begin{figure}
\centering
\includegraphics[width=\columnwidth]{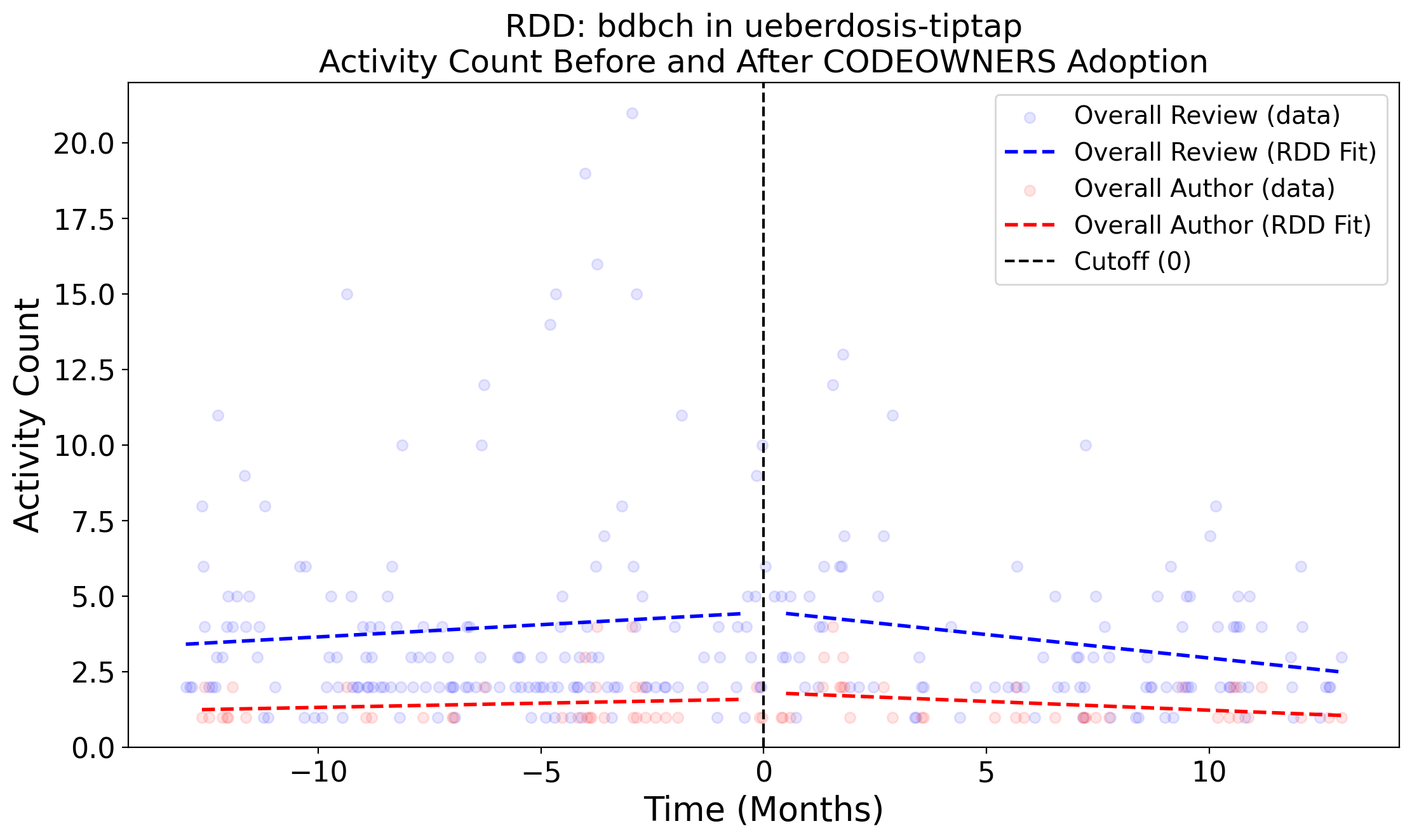}
\caption{Example of developer workload (authoring, reviewing activities) before / after CODEOWNERS adoption.}
\label{fig:rdd_devs_workload}
\vspace{-1.5em}
\end{figure}

An illustrative example is shown in Figure~\ref{fig:rdd_devs_workload}, depicting the activity patterns of developer \texttt{bdbch} in repository \texttt{ueberdosis/tiptap}.
Following the introduction of CODEOWNERS, both authoring and review activities declined noticeably, \rtwoone[4]{suggesting a more selective engagement by this owner after adoption.}

To generalize beyond individual examples, we grouped developers into two mutually exclusive sets: CODEOWNERS (n = 704) and contribution-based owners (n = 3,121).
Developers appearing in both groups were excluded from the contribution-based owners set to prevent overlap\rtwoone[1]{; owner names are matched to GitHub logins exactly and case-insensitively}.
\rtwoone[1]{Designated owners are observed in 139 of the 222 repositories; the other 83 name only teams, or no parsable owner, in their CODEOWNERS file and contribute contribution-based owners only. Table~\ref{table:prs_combined_metrics_trend_post_adoption_codeowners_detailed} reports the estimates.}


\rtwoone[1]{At the cutoff, designated owners' activity rises in all three categories (review $p=0.017$, decision $p=0.001$, authoring $p=0.011$). Contribution-based owners also rise in decision ($p=0.022$) and authoring ($p=0.046$) activity, but not reliably in review ($p=0.15$). No pre-slope and no post-slope differs from zero ($p\geq0.32$), nor does any absolute post-adoption slope $\beta_{1}+\beta_{3}$ ($p>0.2$), so we claim no acceleration or deceleration for either group. The last two columns of Table~\ref{table:prs_combined_metrics_trend_post_adoption_codeowners_detailed} put the jumps in context. Over the year after adoption, designated owners' review actions rose by 61\% and their approvals and merges by 73\%, against a 27\% rise in the pull requests of their repositories: 26\% more review work and 36\% more decision work per pull request than before. Their authoring rose by 36\%, in step with pull-request volume. Contribution-based owners did 20\% less review work per pull request than before, 11\% less decision work, and 15\% less authoring. Within the same 139 repositories the contrast is sharpest: contribution-based owners' review actions there fell by 8\% while designated owners' rose by 61\%. These ratios are descriptive context for the regression estimates: they pool each group's actions across its repositories and so weight repositories by their activity.}

\rtwoone[4]{Appendix~\ref{app:rq4_participants} also tallies the actions of every pull-request participant in the 139 repositories, including developers in neither owner group, to tell whether the owners' additional work is taken over from other developers or added to the repository's total. Human review actions per pull request were approximately unchanged after adoption (ratio 1.02), while designated owners' share of them rose from 20\% to 25\% and contribution-based owners' share fell from 45\% to 32\%; the share of developers in neither group rose from 35\% to 43\%. Decision actions per pull request rose by 7\%, with designated owners' share up from 29\% to 36\%. Authoring per pull request was approximately unchanged (ratio 0.98); designated owners' share of it rose from 19\% to 21\% and contribution-based owners' fell from 46\% to 37\%. The additional review work of designated owners is thus a redistribution of an approximately unchanged number of review actions per pull request toward the declared owners, whereas their additional approvals and merges partly add to the decisions a pull request receives.}

\begin{figure}
\centering
\includegraphics[width=\columnwidth]{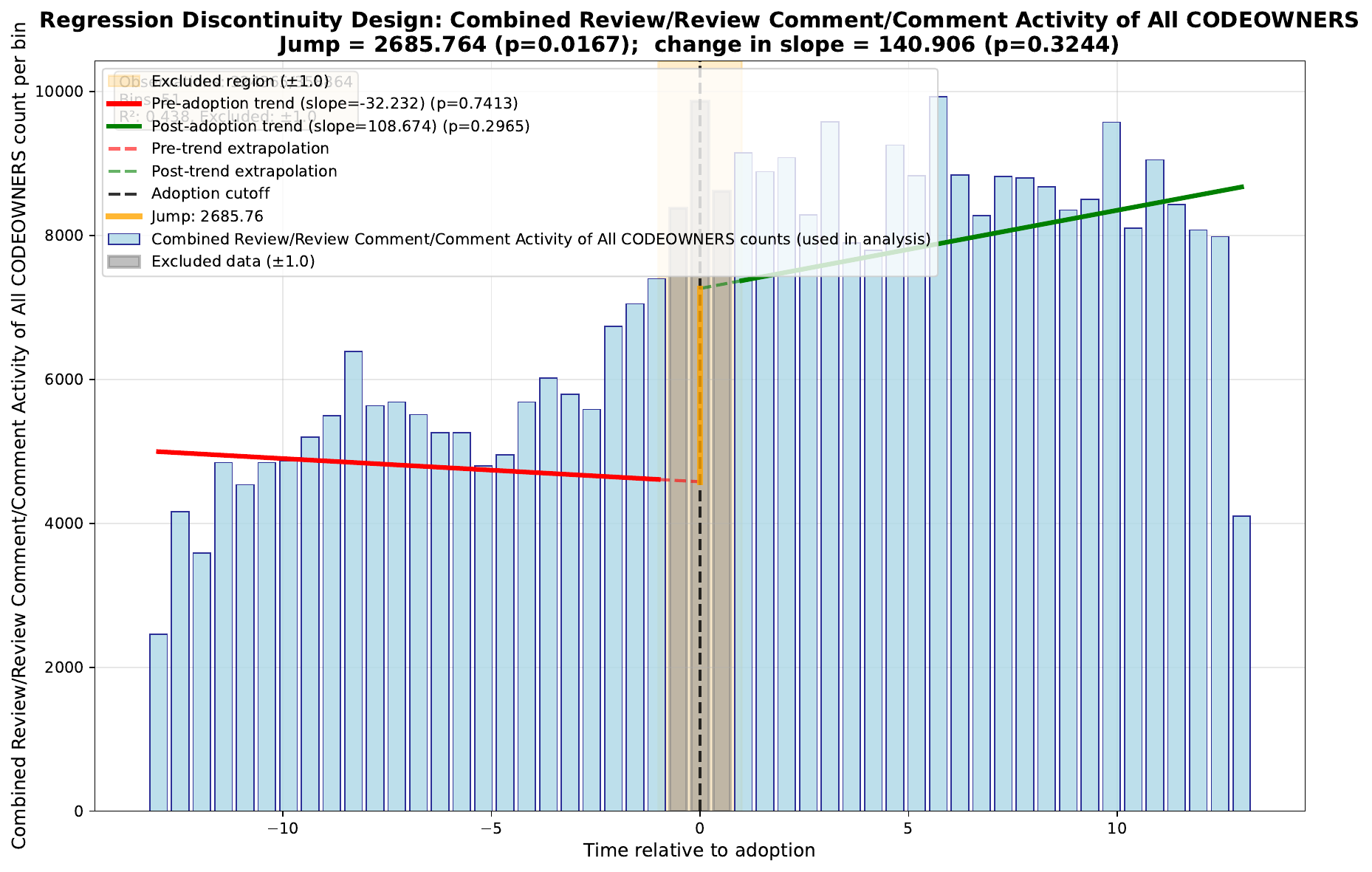}
\caption{\rtwoone[1]{Review activity of designated owners per half-month bin around adoption, with the fitted pre- and post-adoption segments. Cutoff $+2{,}686$ actions per bin ($p=0.017$); pre-slope $-32$ per month ($p=0.74$). The fitted post-adoption segment has slope $+109$ per month ($p=0.30$); the change in slope at adoption, the post-slope quantity of Table~\ref{table:prs_combined_metrics_trend_post_adoption_codeowners_detailed}, is $+141$ ($p=0.32$).}}
\label{fig:rdd_combined_review__codeowners_dev_workload}
\end{figure}

\rtwoone[1]{Figure~\ref{fig:rdd_combined_review__codeowners_dev_workload} shows designated owners' review activity: the bins jump at adoption and stay above the pre-adoption level, while neither fitted slope differs reliably from zero.}

\rtwoone[4]{Together, these findings show that designated owners take on more review and decision work at CODEOWNERS adoption, and that their authoring rises as well. Counting every pull-request participant, the review part of this is a redistribution rather than added review (Appendix~\ref{app:rq4_participants}; caveats in Section~\ref{sec:threats}).}
\revtwo[M5]{We note that an increase in owners' review workload after adoption is, in itself, the behavior CODEOWNERS is designed to produce---it automatically routes review requests to designated owners---so this result also serves as a sanity check that the mechanism operates as intended. \rtwoone[1]{The less obvious result is that owners' authoring keeps pace with pull-request volume rather than giving way to review work, and that owners do more review and decision work per pull request after adoption while contribution-based owners do less review work per pull request.}}

\summarybox{RQ4}{Impact of CODEOWNERS on Developer Workloads}{
\rtwoone[4]{At CODEOWNERS adoption, designated owners' review, decision, and authoring activity rises at the cutoff, most strongly for approvals and merges. Over the following year they do 26\% more review work and 36\% more decision work per pull request than before, their authoring keeps pace with pull-request volume, and contribution-based owners do less review work per pull request than before. Neither group shows a statistically significant pre-slope or post-slope. Counting every pull-request participant, the number of review actions per pull request is approximately unchanged while designated owners' share of it rises and contribution-based owners' share falls: CODEOWNERS redistributes review work toward the declared owners rather than adding review, and adds decision actions (approvals and merges) per pull request.}}

\vspace{-2em}
\section{Implications}

In this section, we interpret our findings in light of practical applications and theoretical implications.
Across all stakeholder groups, our results show that GitHub's CODEOWNERS feature—though sparsely adopted—represents a shift from implicit, contribution-based ownership to explicit, policy-driven governance. \revtwo[M7]{That CODEOWNERS embodies explicit, policy-driven governance is true by construction rather than a finding of our study; what our results add is empirical evidence on how that built-in capability is actually \emph{used}---which projects adopt it, how completely, and how it reshapes review behavior.}
By codifying responsibility in machine-readable configuration files, it operationalizes accountability and compliance within the repository itself, strengthening both governance and supply-chain resilience.
Our evidence suggests that CODEOWNERS formalizes existing ownership behaviors, \rtwoone[4]{raises the review and decision activity of designated owners}, and contributes to more structured and \revtwo[M7]{auditable collaboration processes---that is, processes in which the reviewers responsible for a change are explicitly recorded and can be held accountable after the fact}.

\subsection{Implications for Repository Maintainers}

Only 6.75\% of the top-starred repositories on GitHub currently use CODEOWNERS (preliminaries).
This limited adoption may reflect low awareness or concerns about the maintenance overhead of this alternative approach.
\revtwo[S2]{Low adoption, however, should not be read as evidence that the mechanism is under-adopted across all projects. Our study does not establish an appropriate or optimal adoption level, and we do not claim that every project should adopt CODEOWNERS. The claim our evidence supports is narrower, concerning where adoption is most beneficial. Projects that adopt CODEOWNERS are already the more active ones, with a median of 4,488 commits and 204 contributors against 1,789 and 135 for non-adopters (Section~\ref{sec:prelim}), \rtwoone[1]{and it is in these projects that, after adoption, the trends in merge time and review effort more often bent downward than upward relative to their earlier course (RQ3).} Adherence follows a different axis: it is highest where coverage is broad, and broad coverage is more common in the smaller projects, since the largest ones more often protect only a few critical files (Table~\ref{table:coverage_group_characteristics}). Our findings therefore motivate adoption most directly for active projects, without implying that those same projects are the ones that follow their ownership rules most closely.}

A key implication for maintainers is that \revthree[11:16]{CODEOWNERS adoption is associated with} \rtwoone[4]{a rise in the activity of designated owners, most clearly in review and decision work}, which implies that the ownership roles may change.
\rtwoone[2]{As RQ1 shows that designated owners take part in a share of the pull requests that touch their files, CODEOWNERS could be utilized by maintainers that would like to reorganize their teams and make review responsibility explicit.}

\revone[1.II]{These observations carry concrete guidance for project management. \rtwoone[1]{The active projects in which merge time and review effort more often decelerated after adoption (RQ3)} are also those whose maintainers stand to benefit most from declaring ownership over their most consequential files, rather than leaving review routing to informal convention. The separation we observe between declared owners and the most frequent committers (RQ2) means ownership can be assigned to domain experts or release managers who are not necessarily the busiest contributors, letting teams place review authority where responsibility lies rather than where commit volume happens to concentrate. \rtwoone[5]{Because adoption raises the review and decision activity of designated owners, who then do more of this work per pull request (RQ4), managers should anticipate this change when planning reviewer capacity. We also observe that ownership is often narrow: the median repository names a single individual owner, and 40\% of CODEOWNERS files contain a single rule (Section~\ref{sec:prelim}). We did not measure review bottlenecks, bus-factor risk, or the consequences of narrow versus broad assignments, so the following is a plausible implication rather than an empirical finding. It also assumes that merging depends on the owner's review or approval, which we did not establish and which holds only where the repository requires owner approval (Section~\ref{sec:threats}). Under that assumption, where one owner is responsible for files that most pull requests touch, and that owner's review work grows with adoption, the owner's availability would bound the review throughput of the project, and broader assignments or periodic rotation would limit this dependence.}}

\subsection{Implications for Repository Contributors}

For contributors, CODEOWNERS transforms informal norms of participation into explicit review and approval pathways.
Our analysis shows that designated owners are often not among the top committers (preliminaries), reflecting a decoupling between development effort and decision authority.
This separation allows projects to distribute review responsibilities to domain experts, compliance officers, or maintainers, reinforcing governance without centralizing control in a few prolific developers.

\rtwoone[4]{The rise in designated owners' review and decision activity at adoption (RQ4) suggests that oversight concentrates on the declared owners.}
\rtwoone[4]{This concentration} improves auditability and reduces the risk of unreviewed changes, but it also introduces new coordination challenges and potential bottlenecks if ownership assignments are too narrow.
Periodic reassessment or rotation of ownership responsibilities can mitigate review fatigue and prevent the formation of gatekeeping hierarchies, maintaining both accountability and inclusivity.

\subsection{Implications for Researchers}

From a research perspective, CODEOWNERS provides a natural experiment in software governance at scale.
It enables the study of how repositories formalize responsibility through platform-level mechanisms rather than social conventions alone.
Our findings highlight three key theoretical implications:

\begin{itemize}
\item \textbf{Governance Mechanisms:} CODEOWNERS exemplifies a transition from emergent, contribution-based control to explicit, policy-driven governance \rtwoone[2]{in which designated owners observably take part in review (RQ1)}.
This evolution invites research on how digital platforms embed organizational rules into development infrastructure and how such embedded governance affects transparency, autonomy, and control.
\item \textbf{Security and Supply-Chain Resilience:} The concentration of ownership rules on configuration and dependency files points to CODEOWNERS as a soft security control.
Future work could examine how ownership policies interact with vulnerability management, dependency updates, and build integrity within continuous integration pipelines.
\item \textbf{Formalization of Implicit Ownership:} Our evidence that CODEOWNERS and contribution-based owners behave similarly (RQ2) suggests that the feature often formalizes preexisting social structures rather than replacing them.
This raises deeper questions about whether codifying responsibility strengthens accountability or merely reifies existing hierarchies.
\end{itemize}

By combining repository analytics with qualitative inquiry, future research can unpack how explicit ownership mechanisms reshape accountability, productivity, and trust in open-source ecosystems.
\section{Threats to Validity}\label{sec:threats}


\textbf{Internal Validity.} A possible threat to internal validity lies in our reliance on observable GitHub metrics—such as pull requests, commits, and CODEOWNERS files—which may be affected by repository-specific configurations, organizational policies, or external factors.
For example, some projects may use custom review workflows or external CI/CD tools that are not visible through GitHub's API.
We reduced this risk by using multiple indicators (commits, comments, and reviews) rather than a single measure.
Another limitation is the incomplete visibility of team-based ownership: GitHub restricts access to the membership of private teams, preventing a full assessment of their involvement.
\rtwoone[2]{Team ownership is common in our PR-level data: two thirds of the eligible pull requests under a repository's latest CODEOWNERS version touch only team-owned files. For these pull requests individual-owner participation cannot be observed and is recorded as absent, so our adherence rates understate participation, most strongly in the high-coverage band.}
\rtwoone[3]{A further internal-validity concern is the one-PR-per-repository design of the RQ1 coverage comparison. Owner involvement in a single pull request depends on its size, files, author, and review complexity, so the per-stage percentages carry sampling variability that one draw cannot reveal. A check over all eligible pull requests of the 222 repositories mined for RQ2--RQ4, under the same eligibility rule, gives the same direction for the merge-stage difference but a gap that is not statistically significant (per-repository owner-merge rates, Cliff's $\delta=0.04$, $p=0.66$). We therefore read the merge-stage result as a tendency rather than a reliable difference.}
\revone[3.III]{A further consideration concerns our RQ1 comparison of adherence between low-coverage (0--10\%) and high-coverage (90--100\%) repositories, since higher CODEOWNERS coverage mechanically increases the chance that any given pull request modifies an owned file. We examined this directly in Section~\ref{sec:findings}: the raw cross-coverage differences are largely a structural coverage effect, and once we condition on pull requests that modify an owned file, owner involvement is similar across coverage levels (Cohen's $h\leq0.13$). We therefore read the unconditional cross-coverage differences as predominantly coverage-induced rather than behavioral.}
\revone[3.IX]{One aspect of our observation window warrants caution: pull-request volume rises over the study period in many repositories, so the post-adoption increase in owners' absolute review activity (Figure~\ref{fig:rdd_combined_review__codeowners_dev_workload}) partly co-occurs with this growth rather than reflecting more intensive per-owner engagement alone. \rtwoone[1]{The RQ4 result is a jump at adoption rather than a gradual rise, and designated owners do more review and decision work per pull request after adoption while contribution-based owners do less review work per pull request (Table~\ref{table:prs_combined_metrics_trend_post_adoption_codeowners_detailed}). Our design cannot, however, separate the part of the jump due to review routing from the part due to the activity that accompanies adoption.}}

\textbf{External Validity.} Our dataset focuses on the 10,000 most-starred GitHub repositories, which are generally large, active, and well-maintained.
As a result, the findings may not generalize directly to smaller or less active repositories, or to private repositories that operate under different review policies.
Future work could expand on our results by examining less popular or organizational repositories to validate whether similar adoption patterns and process changes emerge at smaller scales.
\revthree[1:23]{We further note that GitHub is not used exclusively by open-source projects: many proprietary and private repositories are hosted on the platform and may use CODEOWNERS under different review policies and incentives. Our public-repository focus does not capture this proprietary usage, which remains an avenue for future work.}
\revthree[MAJ1]{Relatedly, our binary framing (CODEOWNERS vs. no CODEOWNERS) does not capture other explicit ownership mechanisms---such as \texttt{OWNERS}/\texttt{MAINTAINERS} files or ownership recorded informally in documentation---so repositories we treat as lacking explicit ownership may in fact coordinate ownership through means our data does not reveal.}
\revone[1.II]{Beyond this, our analysis is confined to GitHub, whereas CODEOWNERS-style features are also available on other platforms such as GitLab; adoption and adherence patterns on those platforms may differ, so extending this study across hosts is an important direction for future work.}
\revone[2.III]{Finally, because data collection concluded on 9 October 2024, repositories that adopted CODEOWNERS shortly before this date contribute a shorter post-adoption window; while our within-repository design mitigates calendar-time bias, the recency of the cutoff limits how much long-run post-adoption behavior we can observe for late adopters.}

\textbf{Construct Validity.} Construct validity concerns stem from how key variables—particularly adherence—were operationalized.
We defined adherence as the extent to which designated code owners participated in PR authoring, reviewing, approving, or merging.
While this provides a measure of observable owner participation, it may not fully capture informal ownership behaviors, such as off-platform discussions or delegated reviews.
To reduce ambiguity, we derived our definitions from prior literature on code ownership~\cite{foucalEtAll2014,Rahman2011,Thongtanunam2016} and validated them against multiple participation types (authoring, reviewing, decision-making).
\rtwoone[2]{A more specific concern is that our adherence measure records whether a designated owner took part in a pull request, not whether the repository required that participation. Owner approval becomes mandatory only when a branch protection rule or a ruleset with the require-code-owner-review option is enabled, and GitHub does not expose the history of these settings: rulesets are public but date from 2023 and show only the current state, classic branch protection is visible only to administrators, and the audit log is not accessible to third parties. Our adherence rates therefore measure observable owner participation and should not be read as compliance with an enforced policy. For reference, of the 120 repositories with at least ten eligible pull requests naming an individual owner, 15 show an owner's approval on at least 90\% of them, and the median repository shows one on 41\%.}
\revone[3.VI]{A further consideration concerns how contribution-based (git~blame) owners are resolved to GitHub identities. Using GitHub's commit-to-account linkage, approximately 90\% of contribution-based owners resolve to a GitHub username, while the remainder are retained under their commit-author name. This residual arises when GitHub cannot associate a commit with an account, either because the account was deleted, or because the commit was authored with an email not registered to a GitHub account. In such cases no username is available to retrieve, so our resolution is as complete as the platform's own linkage permits. Because these contributors are retained rather than discarded, our identification of contribution-based ownership is unaffected. The residual limitation is that name-only contributors cannot be matched to pull-request participants (who are identified by GitHub login), so any review or authoring activity they perform under a linked account may be undercounted; this could slightly attenuate the behavioral comparison in RQ2. These unmapped cases correspond to commits that GitHub cannot associate with any account, rather than to a particular class of developer, so we do not expect them to skew the contribution-based owner sample toward a specific developer type. Attributes such as account age or organization affiliation cannot be compared between mapped and unmapped contributors, since no account is available to read them from for the unmapped ones. Because mapping succeeds for roughly 90\% of contribution-based owners and unmapped owners are retained rather than discarded, restricting attention to high-mapping repositories would exclude only the small minority with substantial unmapped fractions, so the small unmapped tail is unlikely to change the behavioral similarity we report.}
\revtwo[S6]{A final construct consideration is the granularity of our activity measures. They are defined at the pull-request level, and a single pull request often modifies both CODEOWNERS-protected and unprotected files. We record which modified files are protected, but reviews, comments, and approvals are attributed to the whole pull request rather than to individual files or regions of the codebase. We therefore cannot determine whether the increase in owners' review activity within protected areas is offset by reduced activity elsewhere, or whether contribution shifts between protected and unprotected parts of the codebase after adoption. Answering this would require attributing review effort at the file or change-hunk level and tracking it separately across regions over time, which we leave to future work.}

\textbf{Conclusion Validity.} Our conclusions rest on Mann-Whitney U, Cliff's $\delta$, and RDD analysis. Nonetheless, unobserved variables—such as internal governance policies, contributor motivation, or organizational context—may influence both CODEOWNERS adoption and adherence.
While these factors are difficult to quantify at scale, the consistency of directional trends across repositories and metrics provides confidence in the reliability of our findings.
\revthree[MAJ2]{A further concern is broad, calendar-time change in the software ecosystem across the study period, including the wider availability of AI coding assistants, which could independently shift merge times and review effort. Our design offers partial protection: because each repository is anchored on its own adoption date rather than on wall-clock time, and adoption dates are spread across several years, a change tied to a fixed calendar date does not coincide with the relative-time cutoff and cannot by itself produce the discontinuity we estimate. It could still bias the pre- and post-adoption slopes if adoption timing correlates with calendar time. A direct test would track the same metrics in our non-adopting baseline repositories over the same calendar period, checking whether the post-adoption \rtwoone[1]{decelerations} coincide with an ecosystem-wide secular trend rather than with adoption itself; because these repositories have no adoption event, this is a calendar-time trend comparison rather than a discontinuity analysis. We identified this baseline set only at the repository level (the 675 non-adopting repositories used in our preliminary characterization) and did not mine their pull-request-level activity; the calendar-time comparison would require that additional pull-request mining, which we therefore leave to future work.}
\rtwoone[4]{The RQ4 models fit the two owner groups only, so Appendix~\ref{app:rq4_participants} tallies every pull-request participant to tell work taken over from other developers from work added to the repository's total. That tally is descriptive: it weights repositories by their activity; developers in neither owner group are a heterogeneous set that also absorbs contribution-based owners whose git-blame identity could not be mapped to a GitHub account; and bots are separated by the \texttt{[bot]} suffix of their login only. The redistribution we report is one of observed pull-request actions, not of effort in hours.}
\rtwoone[1]{Finally, both temporal analyses classify the change in slope at adoption ($\beta_{3}$), which can differ in sign from the absolute post-adoption slope when a metric was already trending, so a Decelerate label in RQ3 means that the trend bent downward at adoption, not that the metric fell afterwards. In RQ4 neither the pre-slope nor the post-slope is statistically significant, so we claim no change in trend; the group-level fits also pool each group's repositories into one series of a few dozen bins and do not account for repositories entering or leaving the window. The per-pull-request ratios of Table~\ref{table:prs_combined_metrics_trend_post_adoption_codeowners_detailed} are likewise descriptive and weight repositories by their activity; the statistical conclusions of RQ4 rest on the regression estimates alone.}

\vspace{-1em}
\section{Related Work}

Prior research has investigated code ownership from multiple perspectives, including its effects on software quality, authorship attribution, team dynamics, and project governance. Our work extends this body of knowledge by providing the first large-scale empirical analysis of explicit, configuration-based ownership through GitHub’s CODEOWNERS mechanism—contrasting it with contribution-based definitions used in earlier studies.

\textbf{Code Ownership and Authorship in Software Projects.} Code ownership has long been recognized as a determinant of software quality and accountability.
Bird et al.~\cite{bird2011don} and Avelino et al.~\cite{Avelino2019} showed that concentrated ownership is associated with lower defect rates and improved project coordination.
Qu et al.~\cite{Qu2021} demonstrated that incorporating developer identity information enhances bug prediction, while Bird et al.~\cite{Bird2009} linked socio-technical ownership networks to software failure prediction.
Rahman and Devanbu~\cite{Rahman2011} further established that author experience correlates with fewer defects.
Kola-Olawuyi et al.~\cite{Kola-Olawuyi2024} and Koana et al.~\cite{Koana2024,Koana2023} examined ownership of DevOps artifacts and team-level accountability, respectively, emphasizing ownership’s organizational dimension.
Businge et al.~\cite{Businge2017} found similar relationships in Android projects.
Earlier conceptual work by Nordberg~\cite{Nordberg2003} and Sedano et al.~\cite{Sedano2016} discussed strategies for managing ownership, and Thongtanunam et al.~\cite{Thongtanunam2016} and Caglayan and Bener~\cite{Caglayan2012} revisited the ownership–quality relationship in modern review and issue-tracking contexts.

Our study differs by shifting from inferred ownership—typically defined via commit shares or contribution percentages—to explicit declarations encoded in configuration files. This distinction allows us to examine how formalized ownership rules affect collaboration and review behavior at scale.

\textbf{Developer Expertise and Code Review.} Prior research has linked reviewer expertise and workload to review quality.
Hajari et al.~\cite{Hajari2024} and Mirsaeedi and Rigby~\cite{Mirsaeedi2020} highlighted the importance of expertise-aware reviewer selection to balance workload and reduce turnover risks.
Herzig and Nagappan~\cite{Herzig2014} studied ownership in testing, and Shridhar et al.~\cite{Shridhar2014} investigated build ownership and its influence on system changes.

While these works infer ownership from historical activity or domain expertise, \rtwoone[4]{our results (RQ4) show how the activity of explicitly assigned CODEOWNERS changes in practice: their review, decision, and authoring activity rises at adoption, and they then do more review and decision work per pull request, whereas contribution-based owners do less review work per pull request.}

\textbf{Authorship Attribution and Source Code Analysis.} Authorship attribution research has focused on tracing developer responsibility and provenance.
Bogomolov et al.~\cite{Bogomolov2021} proposed language-agnostic authorship identification, and Ji et al.~\cite{Ji2007}, Meng et al.~\cite{Meng2013}, and Gong and Zhong~\cite{Gong2021} explored authorship evolution and hidden contributors.
Hattori and Lanza~\cite{Hattori2009} refined ownership analysis through synchronous change histories.

These studies recognize authorship implicitly, for provenance or plagiarism detection, rather than as a declared policy that routes reviews.

\textbf{Team Dynamics and Project Management.} Work on team dynamics has linked ownership to coordination, quality, and security outcomes.
Meneely and Williams~\cite{Meneely2009} confirmed Linus’ Law by showing that collective review improves security; Chung et al.~\cite{Chung2015} related psychological ownership to retention in open-source teams; and Maruping et al.~\cite{Maruping2009} examined collective ownership and coding standards as coordination tools.

Our findings complement these studies by quantifying how CODEOWNERS institutionalizes collective ownership in distributed settings—transforming psychological and cultural notions of “ownership” into explicit, enforceable governance rules.

\textbf{Specialized Studies in Ownership and Development.} Several domain-specific works have examined ownership in specialized contexts:
Datta~\cite{Datta2014} studied task ownership in mobile OS development;
Taivalsaari et al.~\cite{Taivalsaari2014} discussed multi-device ownership and architectural implications;
Ravitch and Liblit~\cite{Ravitch2013} analyzed memory ownership in C libraries;
and Ahmad~\cite{Ahmad2024} examined ownership trade-offs in microservice coordination.

While these works illuminate ownership within particular domains or artifacts, our study provides a cross-cutting empirical perspective on how explicit ownership mechanisms operate across thousands of repositories and affect observable collaboration outcomes.

\textbf{Emerging Perspectives on Governance and Security.} Recent studies and industry reports have begun to examine code ownership from the lens of governance and supply-chain resilience.
A 2024 comparative study of ownership approximation methods showed that commit-based and line-based ownership yield distinct sets of owners and complementary benefits—commit-based measures better reflect software quality, whereas line-based ownership improves accountability~\cite{thongtanunam2024code}.
At the same time, advances in automated and AI-assisted review practices are reshaping how ownership translates into quality assurance: a 2024 field study of an LLM-based code review assistant found improved bug detection and reviewer awareness, albeit with longer pull-request closure times~\cite{cihan2025automated}.
Beyond academia, GitHub has introduced governance mechanisms such as the cleanowners automation to help maintain accurate CODEOWNERS files~\cite{githubKeepingRepository} and the SERVICEOWNERS model to align code ownership with team-level service maintainership~\cite{githubOrganizeThings}.
Industry security practitioners have also emphasized CODEOWNERS as a lightweight yet effective supply-chain control: GitGuardian, for example, recommends adding CODEOWNERS files to ensure that designated reviewers approve changes to critical paths before merge~\cite{gitguardianThreeMechanisms}.
Together, these emerging perspectives situate code ownership not only as a social and quality-related construct but also as a mechanism of platform-level governance and security assurance.

Our work differs from these by moving beyond conceptual or case-specific insights to provide a large-scale empirical analysis of how explicit ownership mechanisms are adopted and practiced across thousands of open-source repositories, quantifying their effects on collaboration, review dynamics, and governance outcomes.

\vspace{-1em}
\section{Conclusion and Future Work}

Explicit code ownership mechanisms such as GitHub's CODEOWNERS feature enable teams to assign clear responsibility for critical repository components, formalizing review and approval workflows.
Our large-scale analysis of top GitHub repositories shows that CODEOWNERS is selectively adopted and most often applied to configuration, workflow, and documentation files—areas central to project stability rather than source code alone.
We also find that designated owners are rarely the most active contributors, reflecting a separation between development and review roles, \rtwoone[3]{and that owner participation differs most at the merge stage between repositories with low and high coverage, a difference we read as a tendency given the single-PR design, while review and approval participation is similar across the two.}
While CODEOWNERS tends to reinforce structured review and accountability, its broader effectiveness depends on how projects distribute and enact ownership responsibilities.

Future work will explore how explicit ownership can help detect or mitigate security and quality risks, what organizational factors influence adherence, and how ownership practices evolve with automation and AI-assisted review.
\revtwo[S2]{Future studies could also look beyond repository-level activity and size to the problem domains and functional needs that drive adoption. Characterizing which kinds of projects take up CODEOWNERS, and why, may reveal whether adoption concentrates in particular areas, such as security-sensitive or infrastructure-heavy software, adding domain-level depth that our preliminary characterization does not capture.}
\rtwothree[1]{A staggered difference-in-differences comparison against the non-adopting baseline repositories, once their pull requests are mined, is a natural next step for the temporal analyses.}
Together, these directions can deepen our understanding of ownership as a foundation for trustworthy and sustainable open-source development.

\section*{Acknowledgments}

This research was supported by a Tier 1 grant funded by the Ministry of Education in Singapore (22-SIS-SMU-099 [C220/MSS23C017]).



\appendices
\section{Threshold Sensitivity Analysis}
\label{app:threshold_sensitivity}
\revthree[Tbl6]{This appendix reports the threshold sensitivity analysis for the RQ3 \rtwoone[1]{post-slope (slope-change, $\beta_{3}$)} classification discussed in Section~\ref{sec:findings}. Table~\ref{table:threshold_sensitivity} records how the Accelerate/Decelerate/Constant classification of each metric changes as the near-zero threshold is scaled from its default ($0.1\times$MAD) up to $10\times$MAD.}

\begin{table*}[!t]
\centering
\caption{Sensitivity of the \rtwoone[1]{post-slope ($\beta_{post}=\beta_{3}$)} classification to the near-zero threshold. \revthree[Tbl6]{For each metric, cells report the percentage of all analyzed repositories classified as Accelerating (A), Decelerating (D), or Constant (C) as the threshold is scaled from its default ($0.1\times$MAD) to $10\times$MAD. As in Table~\ref{table:developers_workload_metric_trend_post_adoption}, entries within a metric do not sum to 100\% because repositories with a non-significant discontinuity ($\alpha=0.05$, roughly 65\%) are omitted.}}
\label{table:threshold_sensitivity}
\begin{tabular}{lccccccccc}
\toprule
\textbf{Threshold} & \multicolumn{3}{c}{\textbf{Merge Time}} & \multicolumn{3}{c}{\textbf{Number of Comments}} & \multicolumn{3}{c}{\textbf{Number of Reviews}} \\
 & \textbf{A} & \textbf{D} & \textbf{C} & \textbf{A} & \textbf{D} & \textbf{C} & \textbf{A} & \textbf{D} & \textbf{C} \\
\midrule
$0.1\times$MAD (default) & 13.0 & 19.3 & 3.1 & 14.3 & 17.9 & 3.1 & 11.2 & 20.9 & 2.1 \\
$1\times$MAD           & 7.3  & 13.0 & 15.1 & 11.5 & 9.9  & 14.6 & 4.2  & 17.2 & 18.2 \\
$2\times$MAD           & 2.1  & 5.2  & 28.1 & 4.7  & 5.7  & 25.5 & 3.1  & 11.5 & 25.0 \\
$5\times$MAD           & 0.5  & 1.0  & 33.9 & 0.0  & 3.1  & 32.8 & 1.6  & 2.6  & 35.4 \\
$10\times$MAD          & 0.0  & 0.5  & 34.9 & 0.0  & 1.6  & 34.4 & 0.5  & 0.5  & 38.5 \\
\bottomrule
\end{tabular}
\end{table*}

\section{Statistical Tests of Adherence Differences}
\label{app:adherence_tests}
\revtwo[S3]{This appendix reports the reliability of the per-stage adherence differences between coverage groups discussed in Section~\ref{sec:findings} (RQ1). Each repository contributes a binary owner-involvement outcome at each pull-request stage (a code owner performed the action, or not), so we compare the two size-matched groups (81 repositories each) with Fisher's exact test (Table~\ref{table:adherence_tests}). Only the merge-stage difference is statistically significant ($p=0.027$, odds ratio $=2.29$, Cohen's $h=0.38$); the authoring, review, and approval differences are not ($p\geq0.66$). This confirms that broader coverage is associated with more consistent owner \rtwoone[2]{involvement} specifically at the final integration (merge) stage\rtwoone[3]{, a difference that Section~\ref{sec:threats} qualifies in view of the single-PR design}.}

\begin{table*}[!t]
\centering
\caption{Fisher's exact tests of code-owner involvement at each pull-request stage, comparing high-coverage ($\geq$90\%) and low-coverage ($\leq$10\%) repositories on the size-matched samples of 81 repositories per group. \revtwo[S3]{Each percentage is the share of repositories in the group in which a code owner performed the action; \emph{p} is the two-sided Fisher's exact probability and Cohen's \emph{h} the effect size.}}
\label{table:adherence_tests}
\begin{tabular}{lccccc}
\toprule
\textbf{Stage} & \textbf{High} & \textbf{Low} & \textbf{Odds ratio} & \textbf{Fisher's exact \emph{p}} & \textbf{Cohen's \emph{h}} \\
\midrule
Authored & 17\% & 14\% & 1.33 & 0.664 & 0.10 \\
Reviewed & 26\% & 30\% & 0.83 & 0.726 & 0.08 \\
Approved & 26\% & 27\% & 0.94 & 1.000 & 0.03 \\
Merged   & 40\% & 22\% & 2.29 & 0.027 & 0.38 \\
\bottomrule
\end{tabular}
\end{table*}

\section{Coverage-Controlled Owner Involvement}
\label{app:coverage_control}
\revone[3.III]{This appendix supports the analysis in Section~\ref{sec:findings} (RQ1) of whether the cross-coverage adherence differences are behavioral or a mechanical effect of coverage. Table~\ref{table:coverage_control} compares owner involvement between low- and high-coverage repositories over all pull requests (unconditional) and over only pull requests that modify an owned file (conditional). The large unconditional differences shrink to negligible ones once we condition on the pull request touching an owned file, indicating that the cross-coverage gap is driven mainly by how often a pull request has an owner assigned rather than by owner behavior.}

\begin{table*}[!t]
\centering
\caption{Owner involvement by coverage group, unconditional (all pull requests) versus conditional on the pull request modifying an owned file. \revone[3.III]{Coverage strongly determines how often a pull request touches an owned file (11\% of low-coverage versus 93\% of high-coverage pull requests); conditioning on this removes most of the cross-coverage difference. \emph{h} is Cohen's effect size for the low-versus-high contrast. Both groups are restricted to pull requests opened after the repository adopted CODEOWNERS.}}
\label{table:coverage_control}
\begin{tabular}{lcccccc}
\toprule
\textbf{Stage} & \multicolumn{3}{c}{\textbf{Unconditional (all PRs)}} & \multicolumn{3}{c}{\textbf{Conditional (owned-file PRs)}} \\
 & \textbf{Low} & \textbf{High} & \textbf{Cohen's \emph{h}} & \textbf{Low} & \textbf{High} & \textbf{Cohen's \emph{h}} \\
\midrule
Merged   & 2.0\% & 21.4\% & 0.68 & 18.7\% & 23.2\% & 0.11 \\
Reviewed & 2.6\% & 24.4\% & 0.71 & 24.0\% & 26.3\% & 0.05 \\
Approved & 2.0\% & 21.7\% & 0.69 & 18.1\% & 23.4\% & 0.13 \\
Authored & 1.9\% & 14.6\% & 0.51 & 17.2\% & 15.8\% & 0.04 \\
\bottomrule
\end{tabular}
\end{table*}

\section{Actions of All Pull-Request Participants}
\label{app:rq4_participants}

\rtwoone[4]{The RQ4 models fit the two owner groups only. To tell whether the owners' additional work is taken over from other developers or added to the repository's total, we tallied every action in every pull request of the same twelve-month windows in the 139 repositories with designated owners (93{,}059 pull requests before adoption, 118{,}254 after), attributing each review, review comment and comment, each approval and merge, and each pull request's authorship to its login: a designated owner, a contribution-based owner, a developer in neither group, or a bot (login ending in \texttt{[bot]}). Logins are matched exactly and case-insensitively, as in Table~\ref{table:prs_combined_metrics_trend_post_adoption_codeowners_detailed}. Table~\ref{table:rq4_participants} reports each class's actions and its share of the human actions before and after adoption, and for all human actions their number per pull request after adoption relative to before. The tally is descriptive and, like Table~\ref{table:prs_combined_metrics_trend_post_adoption_codeowners_detailed}, weights repositories by their activity.}

\begin{table*}[!t]
\centering
\caption{\rtwoone[4]{Actions of every pull-request participant in the 139 repositories with designated owners, twelve months after versus before adoption. Shares are among human actions; bots are given as a share of all actions. The ratio given for all human actions is their number per pull request after adoption over that before.}}
\label{table:rq4_participants}
\begin{tabular}{lrrrr}
\toprule
\textbf{Actor} & \textbf{Before} & \textbf{After} & \textbf{Share before} & \textbf{Share after} \\
\midrule
\multicolumn{5}{l}{\textbf{Review}} \\
Designated owners & 124{,}451 & 199{,}914 & 20.4\% & 25.3\% \\
Contribution-based owners & 273{,}740 & 251{,}959 & 44.9\% & 31.9\% \\
Developers in neither group & 210{,}876 & 337{,}417 & 34.6\% & 42.7\% \\
All human actions (per PR 1.02) & 609{,}067 & 789{,}290 & &  \\
Bots (share of all actions) & 23{,}834 & 55{,}303 & 3.8\% & 6.5\% \\
\midrule
\multicolumn{5}{l}{\textbf{Decision}} \\
Designated owners & 41{,}031 & 70{,}893 & 28.8\% & 36.4\% \\
Contribution-based owners & 65{,}653 & 65{,}525 & 46.0\% & 33.6\% \\
Developers in neither group & 36{,}003 & 58{,}458 & 25.2\% & 30.0\% \\
All human actions (per PR 1.07) & 142{,}687 & 194{,}876 & &  \\
Bots (share of all actions) & 828 & 2{,}933 & 0.6\% & 1.5\% \\
\midrule
\multicolumn{5}{l}{\textbf{Authoring}} \\
Designated owners & 17{,}912 & 24{,}324 & 19.3\% & 21.1\% \\
Contribution-based owners & 42{,}355 & 42{,}902 & 45.6\% & 37.2\% \\
Developers in neither group & 32{,}523 & 48{,}135 & 35.1\% & 41.7\% \\
All human actions (per PR 0.98) & 92{,}790 & 115{,}361 & &  \\
Bots (share of all actions) & 269 & 2{,}893 & 0.3\% & 2.4\% \\
\bottomrule
\end{tabular}
\end{table*}

\ifCLASSOPTIONcaptionsoff
  \newpage
\fi

\newpage
\bibliographystyle{IEEEtranS}
\bibliography{IEEEabrv,filteredref.bib}

\begin{IEEEbiography}[{\includegraphics[width=1.05in,height=1.3in,keepaspectratio,clip]{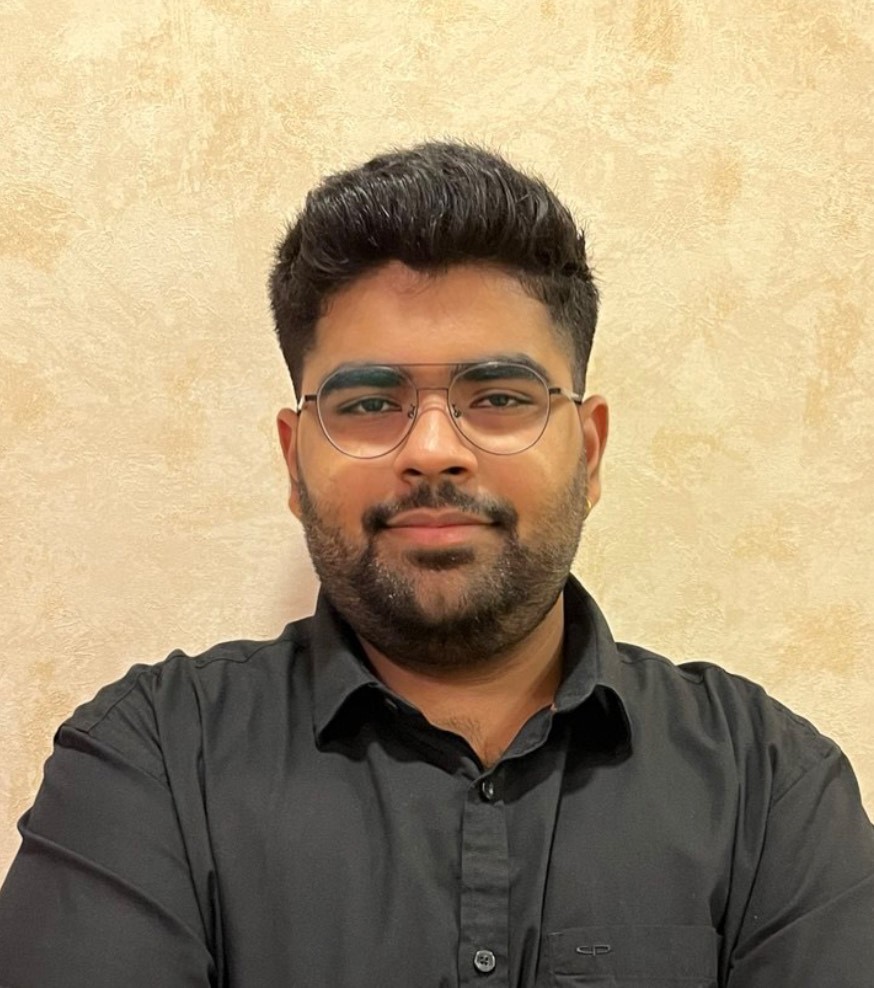}}]{Jai Lal Lulla}
is a Research Engineer at Singapore Management University, working under Professor Christoph Treude. He holds a B.Comp. in Computer Science from the National University of Singapore (NUS).
\end{IEEEbiography}

\begin{IEEEbiography}[{\includegraphics[width=1.05in,height=1.3in,keepaspectratio,clip]{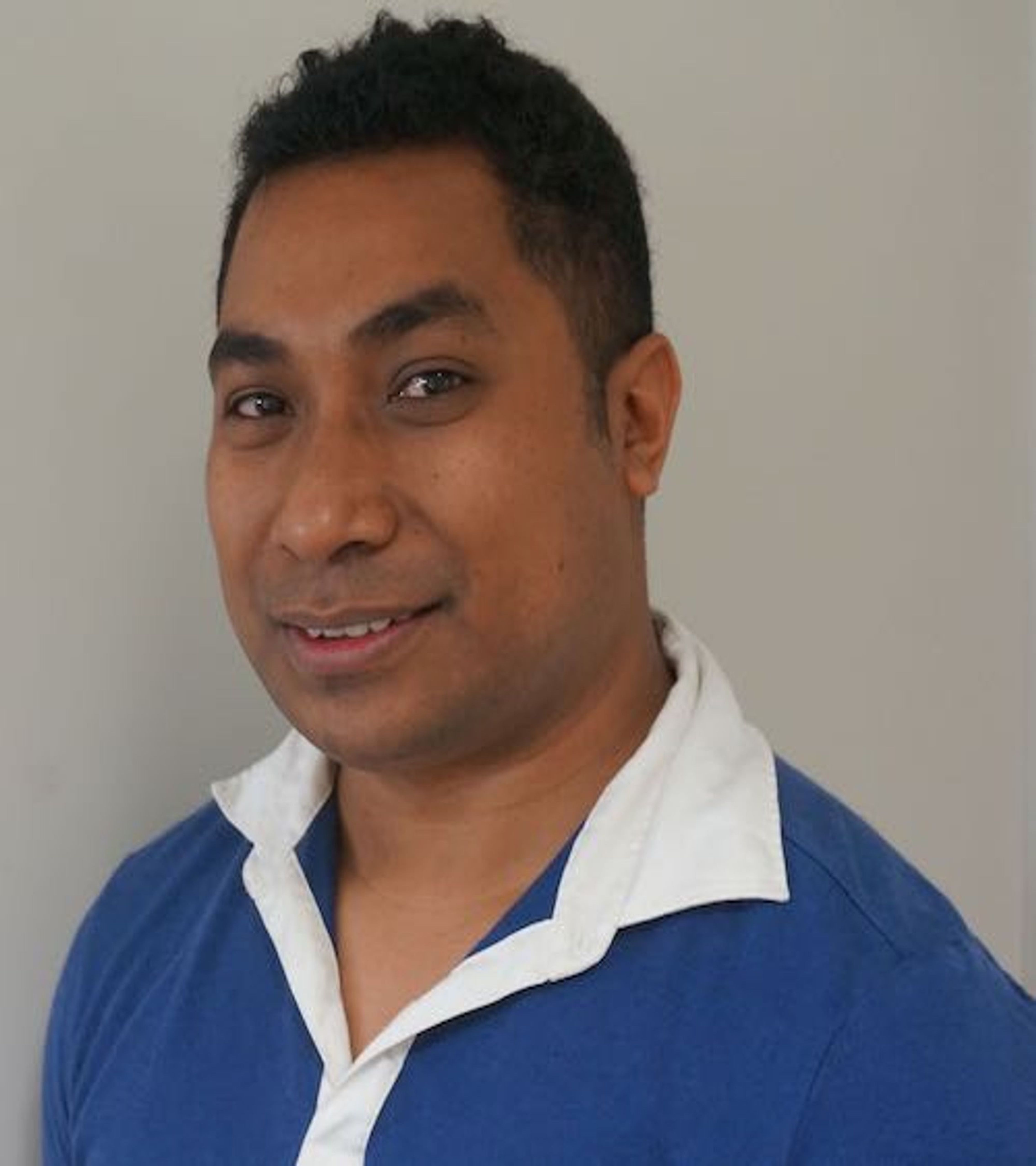}}]{Raula Gaikovina Kula}
is a Professor at The University of Osaka. He received his Ph.D. from Nara Institute of Science and Technology (NAIST) in 2013, later joining as a  Assistant Professor from (2013-2016) at Osaka University. He then continued as a Assistant Professor from (2017), later becoming an Assistant Professor (2017-2023), and a Associate Professor (2023-2024) at NAIST. Kula has published over 150 publications in top Software Engineering venues, collaborating with several researchers from across the globe, and is a member of both Editorial and Steering Committees. Contact him at raula-k@ist.osaka-u.ac.jp
\end{IEEEbiography}

\begin{IEEEbiography}[{\includegraphics[width=1in,height=1.25in,trim=5 0 5 0,clip]{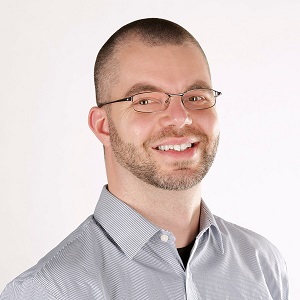}}]{Christoph Treude}
 is an Associate Professor at Singapore Management University (SMU). He serves on the editorial boards of journals including TSE, EMSE, and JSS, and will be PC Co-Chair for FSE 2026. His research goal -- getting information to developers when and where they need it -- is gaining new significance in the AI era. Before joining SMU, Christoph worked in Australia, Brazil, Canada, and Germany. 
\end{IEEEbiography}

\end{document}